\documentclass[aps,prd,twocolumn,superscriptaddress,amsmath,amssymb,nofootinbib,longbibliography]{revtex4-1}

\usepackage[utf8]{inputenc}         % Unicode support
\usepackage[T1]{fontenc}            % Proper font encoding
\usepackage{txfonts}                % Use txfonts package for text and math fonts
\usepackage{graphicx}               % Graphics support
\usepackage{hyperref}               % Hyperlinks
\usepackage{bm}                     % Bold math symbols
\usepackage{amsmath}                % Enhanced math typesetting
\usepackage{amssymb}                % Additional symbols
\usepackage{color}                  % Color support (if needed)

\usepackage{natbib}                 % Bibliography support

\usepackage{array}                  % Custom tables
\usepackage{longtable}              % Tables spanning multiple pages

\newcommand{\be}{\begin{equation}}
\newcommand{\ee}{\end{equation}}
\newcommand{\bs}{\begin{split}}
\newcommand{\es}{\end{split}}

\begin{document}
\title{A Causal Schr\"odinger-Newton Theory that Conserves Energy and Momentum}
\author{Daisuke Miki}
\affiliation{Walter Burke Institute for Theoretical Physics and Theoretical Astrophysics 350-17, California Institute of Technology, Pasadena, CA 91125, USA}
\author{Yanbei Chen}
\affiliation{Walter Burke Institute for Theoretical Physics and Theoretical Astrophysics 350-17, California Institute of Technology, Pasadena, CA 91125, USA}
\author{Yiqiu Ma}
\affiliation{National Gravitational Laboratory, Hubei Key Laboratory of Gravitation and Quantum Physics, School of Physics, Huazhong University of Science and Technology, Wuhan, 430074, China}
\author{Yubao Liu}
\affiliation{State Key Laboratory of Low Dimensional Quantum Physics,
Department of Physics, Tsinghua University, Beijing, 100084, China}
\author{Selim M. Shahriar}
\affiliation{
Department of Electrical and Computer Engineering, Northwestern University, Evanston, IL 60208}
\affiliation{
Department of Physics and Astronomy, Northwestern University, Evanston, IL 60208}
\begin{abstract}

This paper provides a theoretical justification for using the Causal-Conditional Schr\"odinger-Newton theory (CCSN) as a phenomenological model for classical gravity sourced by quantum matter. In CCSN, the gravitational field is driven not by expectation values in the global quantum state, but by conditional expectation values of the stress-energy tensor evaluated on local states determined by measurement records in the causal past. Because these local states are defined operationally from measurement records, the formulation is independent of any particular interpretation of quantum measurement while leading to unambiguous phenomenological dynamics. We apply this framework to a self-gravitating optomechanical system under continuous measurement and show that the resulting stochastic Schr\"odinger evolution, augmented by a classical gravitational field sourced by local conditional stress-energy expectations, can be made consistent with energy-momentum conservation by introducing a corrector tensor. The corrector restores the conservation law without altering the leading Newtonian predictions of CCSN, contributing only negligible post-Newtonian corrections in current experimental regimes. These results place CCSN on firmer theoretical footing and clarify its role as a viable phenomenological model for ongoing and future tests of classical gravity coupled to quantum matter.

\end{abstract}

\maketitle
\section{Introduction}
 
The Schr\"odinger-Newton (SN) theory for semiclassical gravity
has been proposed on various occasions~\cite{carlip2008quantum,moroz1998spherically,salzman2006possible,yang2013macroscopic,giulini2014centre}.  By taking the expectation
value of the stress-energy tensor to drive the gravitational
field, the SN theory predicts a strong self-gravity effect for
macroscopic objects that is experimentally much more accessible
than gravity-induced entanglement experiments.  On the other hand,
in its original form, since a particle's wavefunction is directly
used to source the gravitational potential, the theory is
{\it nonlinear} in the particle's wavefunction, which, even though proposed by various authors~\cite{weinberg1989precision,weinberg1989testing,bialynicki1976nonlinear}, has been
argued to be ``pathological''~\cite{polchinski1991weinberg,gisin1990weinberg,bahrami2014schrodinger}. 

The pathology of nonlinear quantum mechanics appears during the attempt to describe the  quantum measurement process. In standard/linear quantum mechanics, there is a unique mathematical prediction for measurement outcomes, which can then be subject to different {\it interpretations}. According to the Copenhagen interpretation of quantum mechanics,
when a measurement takes place, the wavefunction collapses
instantaneously on a constant-time slice in the experimenter's
preferred Lorentz frame.  This is a non-local process, yet the
causality of physical effects, namely the statistics of
measurement outcomes, is maintained due to the linearity of
quantum mechanics.  According to the Many-World (or Relative
State) interpretation, the wavefunction never collapses, and the
measurement process is simply a unitary interaction between the
system and the measurement device.  In this way, there is no
non-local collapse process, and causality is maintained in all
interpretations of quantum mechanics.  Indeed, different interpretations of quantum mechanics do not lead to observable effect in standard quantum mechanics. 

When nonlinearity was first introduced, it was formulated in ways that seemed to depend on interpretations. 
Within the
{\it Copenhagen interpretation} the measurement outcome on one
party of an entangled pair affects the wavefunction of the other  party instantaneously
across space-like separations; the nonlinearity then allows the
other party to instantly sense the measurement outcome of the
first party, enabling ``superluminal signaling'' and thereby violating
causality.  This was argued by Gisin~\cite{gisin1990weinberg},
and later by Polchinski~\cite{polchinski1991weinberg}, and has
widely been accepted as the main inconsistency of nonlinear
quantum mechanics.  In the Many-World interpretation, nonlinearity
allows different branches of the Universe's wavefunction, which contain macroscopically
distinguishable outcomes, to interact with each other, leading to a rather serious contradiction (sometimes referred to as an ``Everette Telephone''), as argued by Page and Geilker~\cite{page1981indirect}. Nevertheless, Kaplan and Rajendran argued that causal formulations of nonlinear quantum mechanics exist in the many-world interpretation~\cite{kaplan2022causal}.

The potential association with superluminal signaling and the Everette Telephone, as well as the connection to different interpretations of quantum mechanics, has created a lot of skepticism against nonlinear quantum mechanics and Schr\"odinger-Newton theory.  In this paper, we  will explain in detail that the Causal Conditional formulation of Schr\"odinger-Newton (CCSN) theory~\cite{helou2017extensions,helou2019testing,scully2022semiclassical,liu2023semiclassical,liu2025semiclassical,miki2025role} already avoids superluminal signaling and the Everett telephone by construction, restores causality, and fits into both the Copenhagen and Many-World interpretations. The only unconventional feature of this theory is that how gravity is generated depends on the setup of measuring devices --- which does not have an axiomatic definition within quantum mechanics, but proposals
have been made to address this, either from within quantum mechanics (e.g., Quantum Darwinism~\cite{zurek2009quantum}) or by extending it (e.g., collapse models~\cite{bassi2023collapse}).

In order to restore causality in nonlinear quantum mechanics, Kent broke nonlinear quantum mechanics into two separate parts~\cite{kent2005nonlinearity}.  First, the nonlinearity in
the theory enters only through local parameters of the Hamiltonian; second, the local parameters can only depend on information collected within the past light cone of
the location of the Hamiltonian.

Mathematically, at each point $\mathcal{P}$ in spacetime, results of measurement within its causal past $J^-(\mathcal{P})$ can be used to construct a {\it local pure state} $|\psi(\mathcal{P})\rangle$, which turns out to agree with the Tomonaga-Schwinger state on a hypersurface that approaches the past light cone of $\mathcal{P}$~\cite{tomonaga1946relativistically,schwinger1948quantum}.   Parameters of Hamiltonian at $\mathcal{P}$ is constructed from expectation values taken at $|\psi(\mathcal{P})\rangle$.   When viewed in the Heisenberg Picture, the local pure state $|\psi_H(\mathcal{P})\rangle$ is the projection of the initial state $|\Psi(0)\rangle$ into the simultaneous eigenspace consistent with measurement results obtained within $J^-(\mathcal{P})$.  Expectation values of operators at $\mathcal{P}$, taken at the local pure state $|\psi_H(\mathcal{P})\rangle$, are then used as parameters for the local Hamiltonian. 

A more transparent way to understand the restoration of causality is to view Kent's proposal as to  ``realize nonlinear quantum mechanics'' as a quantum feedback control system within linear quantum mechanics, which in turn has been treated  extensively in quantum control
theory~\cite{wiseman2009quantum,jacobs2014quantum}. The nonlinearity of the resulting theory resides in the nonlinear way in which the conditional state is obtained from the initial state and the measurement outcomes. For systems with spatially separated entanglements and non-trivial propagation delays, tools from Algebraic Quantum Field Theory (AQFT) can be adopted to consistently treat such feedback~\cite{haag1964algebraic,fewster2020quantum}.  Furthermore, since the local pure state depends only on measurement outcomes, this proposal also avoids the Everett Telephone by construction. More specifically, since parameters of the Hamiltonian are defined independently from within each branch, they do not allow communication between different branches.

\begin{figure*}
    \includegraphics[width=0.72\textwidth]{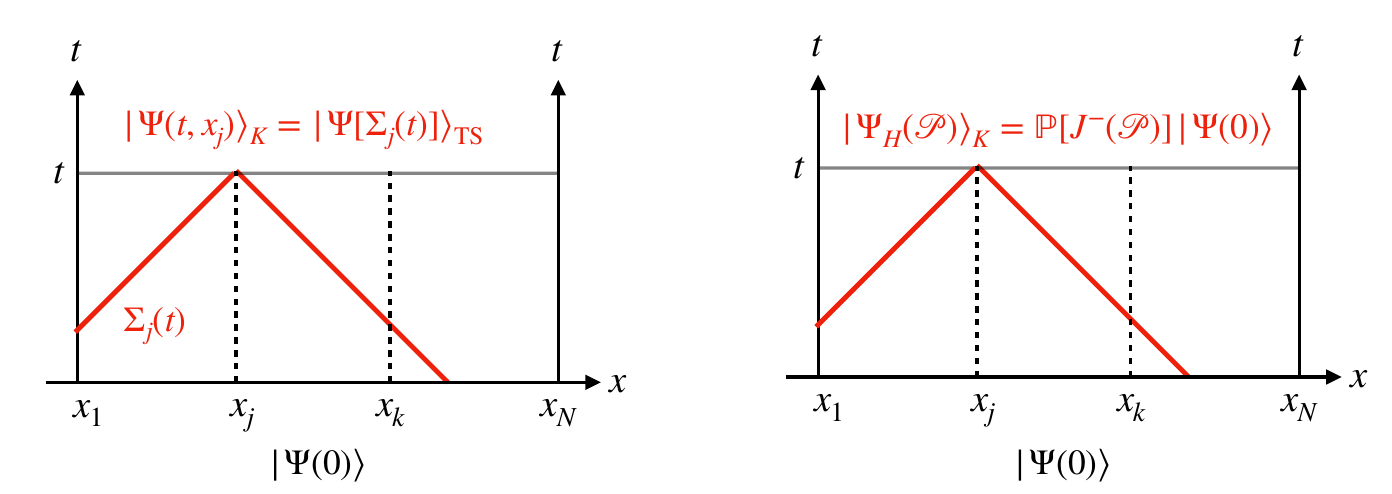}
    \caption{
   Left: local pure states in the Tomonaga-Schwinger Picture;  for each position $x_j$ one defines local pure states along this world line on spatial slices $\Sigma_j(t)$ [see Eq.~\eqref{tomonaga}].  Right: local pure states in the Heisenberg Picture; to obtain $|\Psi_H(\mathcal{P})\rangle$ at space-time event $\mathcal{P}$, projections associated with measurements within the past light cone of $\mathcal{P}$ are included [Eqs.~\eqref{localpureH} and \eqref{pureH2}].  Using expectation values of operators on the local pure state as control parameters allows a causality-preserving implementation of classical interactions within a quantum system.}
\end{figure*}

When Kent's idea is applied to semiclassical gravity, the
gravitational field can be sourced by the expectation value of the
stress-energy tensor operator evaluated at local pure states --- resulting in the CCSN.  Since the CCSN formalism is equivalent to measurement feedback, it can be treated unambiguously following standard/linear quantum mechanics, independent from interpretations.  However, even though  restoring causality, the local state, which
depends on the space-time position, generally does not lead to a
divergence-free stress-energy tensor, with 
\begin{equation}
\label{nonconserv}
    \partial_\mu \langle \psi(x^\rho)|\hat T^{\mu\nu}|\psi(x^\rho)\rangle \neq 0\,,
\end{equation}
as has been pointed out by Wald~\cite{wald2010general}\,(see also for a related constraint in quantum-classical hybrid dynamics\,\cite{Feng2026}).  Such an inconsistency {\it still allows} a Newtonian
calculation, where only the expectation value of the matter
density operator $\hat T^{00}$ is used to compute the Newtonian gravitational
potential; other components of $\hat T^{\mu\nu}$ are not used. It was {\it assumed} that this inconsistency can be
resolved by introducing a corrector field without introducing
significant corrections at the Newtonian level, yet no explicit
construction of such a corrector field was given.  In this paper,
we provide an explicit construction of such a corrector field, and
show that it restores energy-momentum conservation while
introducing only very weak post-Newtonian corrections to gravity --- as
expected.

The experimental signatures of CCSN can differ from those predicted by the standard Schr\"odinger-Newton theory, in which the gravitational field is sourced by the quantum state without conditioning on an actual measurement record. For example, the self-gravity signal proposed for mechanical oscillators in Ref.~\cite{yang2013macroscopic} arises from an SN-induced shift in the frequency governing quantum fluctuations of the oscillator. In CCSN, however, such a signal can be reduced or reshaped because the same measurement used to probe the mechanical motion also provides the classical information used to condition the gravitational source. The observable CCSN signal is therefore generally measurement-protocol dependent and is not determined solely by the underlying Schr\"odinger-Newton coupling.

We would like to point out the connection between the CCSN theory
and other theories of classical gravity, as was discussed in Ref.~\cite{miki2025role}. In Refs.~\cite{bahrami2014schrodinger,nimmrichter2015stochastic,tilloy2016sourcing}, Bahrami {\it et al.}, Nimmrichter and Hornberger, as well as Tilloy and Diosi introduced a stochastic extension of the Schr\"odinger-Newton theory, in which {\it additional} measurement processes are introduced to restore causality.  This also provided a motivation for the Diosi-Penrose collapse model~\cite{penrose1996gravity,diosi1987universal}.  Nimmrichter and Hornberger~\cite{nimmrichter2015stochastic} did imply that other forms of stochastic extensions of the Schr\"odinger-Newton theory are possible; we view CCSN as one such extension, where the stochasticity happens to be arising from the experimentalist's measurement process.  
In the hybrid Classical-Quantum
(CQ) theories formulated by Oppenheim, a series of jump operators are introduced to allow the hybrid evolution of gravity and a quantum system~\cite{oppenheim2023postquantum}. (More generally, Tomizuki and Takeda argued and demonstrated that CQ gravity can be formulated by introducing decoherence from a larger quantum model.)
As shown by Kryhin and Sudhir~\cite{kryhin2025distinguishable}, CQ gravity can naturally incorporate the Diosi-Penrose collapse model as a special case.  Furthermore, as argued by Ref.~\cite{miki2025role}, CQ gravity can be viewed as a measurement-feedback model, and hence also includes the CCSN as
a special case when the only jump operators involved are
observables being monitored by the experimentalist.

The rest of this paper is organized as follows.  In Sec.~\ref{sec:Kent}, we review Kent's argument, as well as the CCSN formulation of Schr\"odinger-Newton theory and how causality is restored. In Sec.~\ref{SN:NoMeasurement}, we review the standard Schr\"odinger-Newton theory without measurements, and show that energy-momentum conservation is satisfied. In particular, we write down the stress-energy tensor operator $\hat T_{\mu\nu}$ for a quantum mass under continuous measurement by a light beam, and explicitly show that $\partial_\mu T^{\mu\nu} = 0$ is satisfied in the Heisenberg picture, and hence $\partial_\mu \langle \hat T^{\mu\nu} \rangle = 0$ is also satisfied in absence of measurements. 
In Sec.~\ref{sec:corrector}, we first discuss how  $\partial_\mu \langle \hat T^{\mu\nu} \rangle $ no longer vanishes when measurement is performed and the local pure state is used to compute the expectation value. We then introduce a corrector field to restore energy-momentum conservation in the CCSN theory, and show that it only introduces very weak post-Newtonian corrections. Finally, we summarize our main conclusions in  Sec.~\ref{sec:conclusion}.

\section{Restoration of Causality: Local Pure State and Stochastic Completion}
\label{sec:Kent}

In this section, for completeness, we provide a brief  review  about the restoration of causality in CCSN, even though this has been explained in Ref.~\cite{miki2025role}.  Kent's restoration of causality is discussed in detail, showing its connection to the Tomonaga-Schwinger picture~\cite{tomonaga1946relativistically,schwinger1948quantum}, as well as state-based quantum feedback control~\cite{wiseman2009quantum,jacobs2014quantum}.  We will then illustrate the restoration of causality using the thought experiment proposed by Bahrami {\it et al.}~\cite{bahrami2014schrodinger}.  We finally provide an additional viewpoint of causality restoration from ``stochastic completion'', and elaborate the connection between our approach and Refs.~\cite{bahrami2014schrodinger,nimmrichter2015stochastic,tilloy2016sourcing}. 

\subsection{The Local Pure State}

Suppose we prepare two objects, $A$ and $B$, in an entangled
state, and we are going to perform a measurement on $A$.  If the
evolution of the wavefunction is nonlinear --- and indeed the
wavefunction ``collapses'' instantly on a constant-time slice in
our Lorentz frame --- then the measurement outcome on $A$ will
affect the wavefunction of $B$ instantaneously.  Kent's fix is to
define, at any space-time event $P$, a local state
$|\psi(P)\rangle$ for the joint system $A$ and $B$ that depends
only on results of measurements in the past light cone of
event $P$.

Let us define $|\psi(P)\rangle$ from within standard quantum
mechanics. Suppose we have a number of localized quantum systems with spatial coordinates $x_1$, $x_2$, \ldots $x_N$, and that the systems have an initial state $|\Psi\rangle$ that might be entangled, and that they interact with each other with causal interactions.  Let us recall that the Tomonaga-Schwinger formulation allows us to define a quantum state $|\Psi(t_1,t_2,\ldots t_N)\rangle_{\rm TS} $, as long as $(t_1,x_1)$, $(t_2,x_2)$, \ldots $(t_N,x_N)$ all lie within a space-like surface~\cite{tomonaga1946relativistically,schwinger1948quantum}. This is possible because operators within space-like separations commute with each other, allowing systems to evolve independently from each other as long as within spatially separated intervals.  For any $x_j$, the Kent local pure state $|\Psi(t,x_j)\rangle_K$ is then given by 
\begin{align}
\label{tomonaga}
    |\Psi(t,x_j)\rangle_{\rm K} & =|\Psi(t-|x_1-x_j|/c+0,\ldots ,t-|x_N-x_j|/c+0)\rangle_{\rm TS}\nonumber\\
    & \equiv |\Psi[\Sigma_j(t)]\rangle_{\rm TS}
\end{align}
In this way, if a measurement is performed on system $k$ at time $t_M$, the local pure state $|\Psi(t,x_j)\rangle_{\rm K}$ only responds to the measurement at $t=t_M + |x_j-x_k|/c+0$; causality is enforced by construction.  Specifically, if we have two objects $A$ and $B$ separated by distance $L$, then a measurement on $A$ performed at $t$ will only affect the local pure state at $B$ after $L/c$. 

The Tomonaga-Schwinger formalism is no longer widely used, as the Heisenberg Picture can be adopted to perform calculations more efficiently.  In fact, the space-time dependence of operators replaces the need to explicitly mention the quantum state, which stays constant except when measurements are performed.   Let us now explicitly treat the measurement process.  Suppose we perform continuous measurements on Heisenberg Operators $\hat Z_j(t)$ of system $j$, and we impose the {\it simultaneous measurability} condition for all the measurements,
\begin{equation}
\label{zzcomm}
[\hat Z_j(t),\hat Z_k(t')]=0\,,\forall t,t'
\end{equation}
which are universally true for observables that do correspond to what are being measured~\cite{braginsky1995quantum}.

Suppose the measurement results for $\hat Z_j(t)$ are $\xi_j(t)$, then the local Heisenberg-Picture pure state is given by 
\begin{align}
\label{localpureH}
    &|\Psi_H(t,x_j)\rangle_K 
    \nonumber\\=& \prod_{k=1}^N \mathbb{P}\left[\hat Z_k(t') =\xi_k(t'): t'<t -|x_j-x_k|/c\right]|\Psi(0)\rangle  
\end{align}
where $\mathbb{P}$ is a projection operator.   This can be concisely written as
\begin{equation}
\label{pureH2}
    |\Psi_H(\mathcal{P})\rangle_K = \mathbb{P}[J^-(\mathcal{P})] |\Psi(0)\rangle
\end{equation}
We can also see explicitly that this local pure state only depends on results of measurement obtained in its causal past. 

An explicitly causal evolution can be obtained if the local pure state is used to generate a parameter for the local Hamiltonian --- a generalized classical force, in a way that will preserve the commutation relations~\eqref{zzcomm}.  Such preservation should be straightforward if the measured quantities are out-going field operators. Evaluating the stress-energy tensor on the local pure state of each system can be used to generate a gravity field which propagates and in turn acts on the entire distributed system --- as we shall discuss later in this paper.

The restoration of causality is most transparent if the Kent model is viewed as a measurement-feedback construction within ordinary linear quantum mechanics. The local pure state at a spacetime point, Eq.~\eqref{localpureH}, is computed from the initial Heisenberg state $|\Psi\rangle_H$, the measurement outcomes in its causal past, and the generalized classical forces previously imposed. Thus Kent’s approach partially linearizes a would-be nonlinear quantum-gravitational dynamics. The residual nonlinearity is the nonlinearity of  conditioning rather than a fundamental nonlinear Schr\"odinger evolution. Expectation values in the conditional state involve normalization by the likelihood of the observed record, and therefore depend on the initial states of both the system and the measuring devices, as well as on the realized record.  In the Appendix, we describe CCSN for two continuously monitored test masses, which has been treated extensively in Refs.~\cite{liu2023semiclassical,liu2025semiclassical,miki2025role}.

Now that classical gravity is formulated in a Lorentz-invariant way that explicitly preserves causality, there are two remaining conceptual issues: (i) the feedback depends only on the past, breaking time-reversal symmetry, and (ii) the evolution at $\mathcal{P}$ depends on {\it results of measurement} within $J^{-}(\mathcal{P})$ --- even though there is no axiom specifying what constitutes a quantum measurement.  The first issue means CCSN depends on the arrow of time --- but this is already a feature of quantum measurement processes.  The second issue leaves a logical hole in the CCSN theory which touches the foundations of quantum mechanics.  We have taken the measurement-based feedback as a phenomenological model for how
classical gravity can be driven by quantum matter.  Our empirical strategy is to consider only systems whose quantum evolution is predominantly determined by the measurements performed by the experimentalist. In such settings, the relevant measurement record is fixed operationally by the experimental setup, including the system-meter coupling, the monitored output channel, and the measured pointer observable or field quadrature. Once these ingredients are specified, the corresponding conditional evolution is fixed in the usual sense of quantum measurement theory.

In CCSN, different experimental measurement choices can lead to different conditional states at the same spacetime point $\mathcal P$, because $J^-(\mathcal P)$ contains different classical records. Since gravitational field is sourced by the conditional expectation value of the stress-energy tensor, these choices can in principle lead to different record-conditioned gravitational fields, distinguishable through correlations between the source measurement record and the gravitational response of a probe. It is in this context that we put the Schr\"odinger-Newton theory to test.

\begin{figure*}[ht]
    \includegraphics[width=0.6\textwidth]{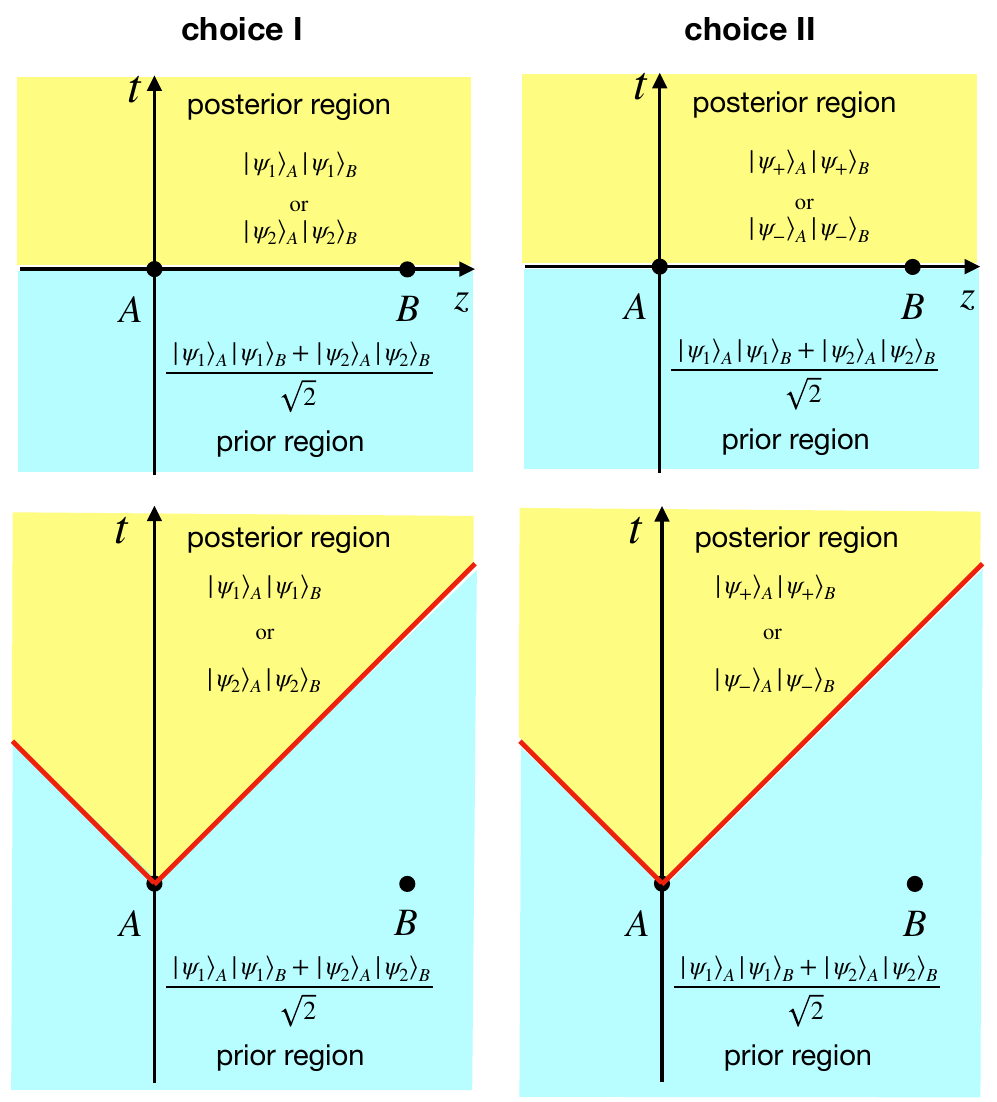}
    \caption{Masses A and B are separated by $L$ along the $z$ direction, with their initial joint quantum state for their $x$ displacements prepared as an entangled state. Upper Panels: super-luminal signaling within the standard
    Schr\"odinger-Newton theory, which uses a wavefunction that
    collapses instantly across the constant-time slice $t=0$ in which
    the measurement is made.   Lower Panels: explicit restoration of
    causality when the result of the measurement is propagated at
    the speed of light across spacetime. In both choices, the local pure state of B used for generating gravity in its vicinity does not collapse due to measurement until the light ray from $A$ reaches $B$, and then $B$ starts to have further evolutions affected by the measurement on $A$. 
    \label{fig:nosignaling}
    }
\end{figure*}

\subsection{Bahrami et al.'s thought experiment}

In order to see how the Kent/CCSN argument resolves causality,  let us  see how Bahrami's thought experiment does not lead
to superluminal signaling in Kent's formulation~\cite{bahrami2014schrodinger}. 

In this thought experiment, one prepares two particles, $A$ and $B$, into an entangled state
\begin{equation}
    |\psi_1 \rangle_A |\psi_1\rangle_B +|\psi_2 \rangle_A |\psi_2\rangle_B 
\end{equation}
where $\psi_1$ and $\psi_2$ are two substantially separated
wavepackets in the $x$ direction. To be concrete, one can write 
\begin{equation}
    \psi_{1,2}(x) =(\sqrt{2\pi }\sigma)^{-1/2} e^{-(x \pm d)^2/(4\sigma^2)}
\end{equation} 
 Next,  particle $B$ travels
toward a large distance $L$ along the $z$ direction away
from $A$.  

Suppose, when $A$ and $B$ are separated by a spatial distance
$L$, $A$ has the choice of performing one of two measurements,
with choice~I projecting into either $|\psi_1\rangle$ or
$|\psi_2\rangle$, and with choice~II projecting into either
$|\psi_+\rangle \equiv (|\psi_1\rangle + |\psi_2\rangle)/\sqrt{2}$
or
$|\psi_-\rangle \equiv (|\psi_1\rangle - |\psi_2\rangle)/\sqrt{2}$. We shall now investigate the consequence of this measurement, both in the standard, instant-state-collapse mode, and in the Kent/CCSN formulation. 

Let us start with the instant collapse mode,  in which 
choice~I will collapse $B$, instantly, to either $|\psi_1\rangle$
or $|\psi_2\rangle$, while choice~II will collapse $B$, instantly,
to either $|\psi_+\rangle$ or $|\psi_-\rangle$, as illustrated in
the upper panels of Fig.~\ref{fig:nosignaling}.  For choice~I,
either $|\psi_1\rangle$ or $|\psi_2\rangle$ will source a
gravitational field centered around $x=d$ or $x=-d$, meaning that
object $B$ will remain at the collapsed location, therefore $x_B$ is either $d$ or $-d$.  For choice~II,
however, $|\psi_+\rangle$ and $|\psi_-\rangle$ will source a
gravitational field that causes $B$ to move toward $x=0$ and oscillate.  Therefore, by observing the motion of $B$, one can instantly determine the choice of measurement performed on $A$, constituting superluminal signaling.  We also note that by ``instant'' we mean
the posterior state changes as we cross a constant-time surface,
which is frame dependent. 

This situation is easily resolved if we adopt the Kent/CCSN formulation.  Using the simplest Tomonaga-Schwinger picture, $|\Psi(B)\rangle_K$ only starts being collapsed at time delay $L/c$ from the measurement at $A$.  Therefore, the
gravitational field at $B$ does not change until this signal
arrives, and no superluminal signaling is possible.  This is
illustrated in the lower panels of Fig.~\ref{fig:nosignaling}.

More specifically,  in Figure~\ref{fig:nosignaling}, in the prior region, we are not affected by  choices I and II. In particular, from $t=0$ until $t = L/c$, the local pure state at $B$ is still the initial state of the system, and  mass B will always have an oscillatory distribution because of the co-existence of $|\psi_1\rangle_B$ and $|\psi_2\rangle_B$ in (different terms of) the wavefunction. According to the Kent theory, the local pure state at $B$ at time $t$ is defined along the $\Sigma_B(t)$ Tomonaga hypersurface, which connects $(z_A,t-L/c)$ and $(z_B,t)$. We then have a joint state of 
\begin{align}
    |\Psi(B,0)\rangle_K=|\Psi[\Sigma_B(0)]\rangle_{\rm TS}=\frac{    |\psi_1\rangle_A |\psi_1\rangle_B +|\psi_2\rangle_A |\psi_2 \rangle_B}{\sqrt{2}}
\end{align}
Here we have further assumed that the state of $A$ does not evolve over time.  Now, with $t$ increasing, measurement does not take place in this slice of $\Sigma_B(t)$ until $t=L/c$.  During this time, $B$ has a Hamiltonian that contains classical gravity accoriding to $|\Psi(B,t)\rangle_K$, and therefore the two packets attract each other. 
We can write 
\begin{align}
    |\Psi(B,t)\rangle_K=\frac{    |\psi_1\rangle_A \hat U_B(t) |\psi_1\rangle_B +|\psi_2\rangle_A \hat U_B(t)  |\psi_2 \rangle_B}{\sqrt{2}}
\end{align}
where $\hat U_B$ is the evolution operator for $B$ under the two-mass gravitational potential.

Along the light cone of the measurement on $A$ (red lines in lower panels of Fig.~\ref{fig:nosignaling}), we will collapse that joint state between $A$ and $B$. 
If choice I is made, depending on the result of $A$, $B$ will be selected as one of the components, 
\begin{equation}
    \hat U_B(t) |\psi_1\rangle_B\,,\quad\mathrm{or}\quad
\hat U_B(t)    |\psi_2\rangle_B
\end{equation}
each a displaced single packet.  
If choice II is made instead, $B$ will be 
\begin{equation}
    \hat U_B(t) |\psi_+\rangle_B\,,\quad\mathrm{or}\quad
\hat U_B(t)    |\psi_-\rangle_B
\end{equation}
which are double packets.  These two scenarios will evolve differently, but this difference only starts at $t = L/c$, which is causally connected to the measurement event on $A$. 

\subsection{the Stochastic Schr\"odinger-Newton Equation (SSNE) and Stochastic Completion}
\label{subsec:SNNE}

Having resolved the issue of causality, for slow-motion sources, let us write down the  Causal conditional formulation of Schr\"odinger-Newton (CCSN) theory for a series of masses under continuous measurement performed by an experimentalist.  This is formulated as a Stochastic Schr\"odinger-Newton Equation satisfied by the {\it conditional state} $|\psi\rangle$ of the system, 
\begin{align}
    \label{eq:full_theory}
d|\psi\rangle &= -\frac{i}{\hbar} \left( \hat H_0 + \hat V_{\rm SN}^j \right) |\psi\rangle dt\nonumber\\
& + \sum_j \frac{\alpha_j (\hat{A}_j - A_{c,j}(t))}{\sqrt{2}} |\psi\rangle dW_j \nonumber\\
&- \sum_j \frac{\alpha_j^2}{4} (\hat{A}_j - A_{c,j}(t))^2 |\psi\rangle dt
\end{align}
where the $\hat V_{\rm SN}^j$ is the Newtonian gravitational potential at the position of the $j$-th mass, {\it evaluated at the conditional state $|\psi\rangle$, with:}
\begin{equation}
    \hat V_{\rm SN}^j = M_j \Phi(\hat X_j) .
\end{equation}
and 
\begin{align}
    \Phi(\mathbf{\hat X}_j) &= -G \int d^3 \mathbf{y} \frac{\langle \psi|\rho(\mathbf{y}-\hat{\mathbf{X}}_j)|\psi\rangle}{|\mathbf{x}-\mathbf{y}|}\,.
\end{align}
Here $\hat A_j$ is the operator being measured on the $j$-th mass with measurement strength $\alpha_j$, and  the $j$-th measurement record $y_j$ satisfies a stochastic differential equation, with 
\begin{equation}
\label{eqyj}
    dy_j =\alpha  A_{c,j}(t) dt + \frac{dW_j}{\sqrt{2}}\,,\quad A_{c,j}(t) = \langle \psi | \hat A_j | \psi \rangle\,.
\end{equation}
where  $A_{c,j}(t)$ is the conditional expectation value of $\hat A_j$ on the conditional state $|\psi\rangle$, and {\it dW } is the Wiener  increment.   Note that these $\hat A_j$ are what the experimentalists are measuring, and can be more general than position operators; there can also be multiple measurements performed at each site.  The CCSN formulation was proposed in Refs.~\cite{helou2017extensions,helou2019testing,scully2022semiclassical}, but these set of Stochastic Sch\"odinger-Newton Equations (SSNE) were first proposed by Liu and Ma in Refs.~\cite{liu2023semiclassical,liu2025semiclassical}.

We note that the dependence of $\Phi(\cdot)$ in the Liu-Ma SSNE on the conditional state $|\psi\rangle$ is a consequence that $|\psi\rangle$ can be computed from the out-going fields $y_j$ and the initial state of the system, as explained, e.g., by Refs.~\cite{PhysRevA.62.012105,wiseman2009quantum,jacobs2014quantum}. In Ref.~\cite{wiseman2009quantum}, systems that have interaction Hamiltonian terms that depend on the conditional state are referred to as {\it state-based quantum feedback control systems} and are discussed at length in Chapter 6.  In Sec.\ 5.3.1 of Ref.~\cite{jacobs2014quantum}, the feedback control Hamiltonian also depends explicitly on the conditional state [see their Eq.~(5.6)].   In other words, the SSNE is an ordinary Stochastic Schr\"odinger Equation (SSE) that describes a state-based quantum feedback control system, and is therefore free from superluminal signaling.  The nonlinearity in this equation originates from the quantum measurement processes that takes place during the system's evolution.

The current form of Eq.~\eqref{eq:full_theory} can also be viewed as an {\it alternative version} of a {\it stochastic extension of Schr\"odinger-Newton theory}, as proposed Refs.~\cite{nimmrichter2015stochastic,tilloy2016sourcing,bahrami2014schrodinger}.  In those works, additional measurements were introduced to provide information to generate classical gravity, leading to additional stochastic terms in the Stochastic Sch\"odinger Equation, which in turn leads to Lindblad terms in the Master Equation. The insight underlying the CCSN is that these additional terms are mathematically unnecessary if measurement records of the experimentalist  is used to generate classical gravity. The corresponding physical difference is that the setup of the measurement influences gravity in a nontrivial way.

\section{Schr\"odinger-Newton Theory Without Measurements}
\label{SN:NoMeasurement}

Before moving on to measurement-induced Schr\"odinger-Newton
theory, let us first review the standard Schr\"odinger-Newton
theory without measurements.  In this case, the wavefunction of
the system is not being collapsed, and therefore the expectation
value of the stress-energy tensor is calculated on the uncollapsed
wavefunction.  In this way, the stress-energy tensor is conserved,
and there is no causality issue.  We will explicitly show this for
a point particle, as well as a matter distribution, in the
slow-motion approximation.  This section also serves as a starting
point for the next section, where the space-time dependence of the
{\it local pure state} will cause corrections.  We will start from
the classical energy-momentum budget of a mass that is being moved
by a light beam, and then move on to the quantum case.  

\subsection{Stress-Energy Tensor Associated with Classical Dynamics 
of a Test Mass Interacting with a Light Beam}
\label{sec:classical_em}

In linearized gravity, we use the trace-reversed metric perturbation
$\bar h^{\alpha\beta}$ to describe the gravitational field. In the 
Lorenz gauge with $\partial_\nu \bar h^{\mu\nu} = 0$, the 
Einstein field equations are given by 
\begin{equation}
    \label{eq:wave_equation}
\square \bar h^{\mu\nu} = -16\pi G {T}^{\mu\nu} 
\end{equation}
with $T^{\mu\nu}$ the stress-energy tensor of the source.  The 
consistency with the Lorenz gauge condition requires 
$\partial_\nu T^{\mu\nu} = 0$, which is the energy-momentum 
conservation condition. In this subsection, let us consider a classical 
test mass that is being moved by a light beam (the EM field), and see how 
the energy-momentum conservation condition is satisfied.

Let us work within one spatial and one temporal dimension for simplicity.
The mass has a matter density of $M \delta (x-X)$, where $X$ is its 
position.  The mass has a momentum density
of $M \dot X\delta (x-X)$, and stress of $M \dot X^2 \delta(x-X)$.
The light beam has an EM (Maxwell) stress of $F \Theta(X-x)$,
corresponding to a momentum flux, with $\Theta$ the Heaviside step
function. The total stress-energy tensor, in the $1+1$ dimensional
case, is given by 
\begin{align}
    T^{00} & = M  \delta (x-X) \,,\nonumber\\
    T^{10}=T^{01} & = M  \dot X \delta (x-X) \,,\nonumber\\
    T^{11} & = M \dot X^2 \delta (x-X) + F(t)\Theta(X(t)-x)\,.
\end{align}
Notice here that we have set $c=1$ for simplicity.  In reality,
even though the EM field has contributions to $T^{00}$ and
$T^{01}$, they are negligible compared to the contribution from
the mass, and therefore we have ignored them.  We can verify that
this stress-energy tensor is conserved. In particular 
\begin{align}
    \label{eq:class:energy_conservation}
    \partial_\nu T^{0\nu} &= \partial_t T^{00} + \partial_x T^{01} 
    \nonumber\\
    & = - M \dot X \delta'(x-X) +M \dot X \delta'(x-X)  =0
\end{align}
which is the energy conservation condition, and 
\begin{align}
    \partial_\nu T^{1\nu} &= \partial_t T^{10} + \partial_x T^{11} 
    \nonumber\\
    & = M\ddot X \delta(x-X) -M \dot X^2 \delta'(x-X) \nonumber\\
    & + M \dot X^2 \delta'(x-X) - F(t) \delta(X-x),
\end{align}
which vanishes if we have $M\ddot X = F(t)$, which is the
momentum conservation condition. From this, we see that
energy-momentum conservation requires us to account for not only
the motion of the test mass but also the system that causes the
motion of the mass --- the EM field in this case --- and the story
is only complete when we account for energy and momentum
conservation of the entire system.  

In a physical situation, the EM fields has to come from somewhere. 
For example, if the EM field acting on our test mass is emitted
directly from a suspended laser device, then the laser device
receives an opposite momentum kick -- a recoil.  If our mass is the end mirror of a LIGO-like interferometer's high-finesse arm cavity, then most of the momentum kick is
imparted onto the cavity's front mirror~\cite{kimble2001conversion,pang2018quantum,pang2019fundamental}.  In both cases, in addition to the local conservation of momentum, we can see that the total momentum that is conserved consists of the momentum of the masses at both ends of the beam --- having ignored field momentum of the beam. This cancellation of recoil 
momenta also manifests in the cancellation 
of the dipole component of the gravitational field associated with the test mass and the recoiling mass --- altogether gravitational radiation
starts at quadrupole order. 

\subsection{Classical Linearized Gravity}

If we restrict to slowly moving sources
dominated by non-gravitational interactions, we can use $T^{00}$
and $T^{0j}$ to generate the leading contributions to the
gravitational field, and therefore we can ignore the contribution
from $T^{ij}$, which is of order $(v/c)^2$ smaller. Since the EM field's contributions to $T^{00}$ and $T^{01}$ are also suppressed by $(v/c)^2$ compared with the mass contribution to $T^{00}$, we can ignore them and focus only on
the matter density and momentum density of the test mass. Using Eq.~\eqref{eq:wave_equation} and
 only keeping the $00$ and $01$ components, this will lead to a metric of 
\begin{equation}
    \label{eq:metric}
    ds^2 = -(1 -2\Phi) dt^2 -4 A_i dx^i dt +(1+2\Phi) \delta_{ij} dx^i dx^j
\end{equation}
with 
\begin{align}
    \label{eq:PhiA}
\Phi(t,\mathbf{x})&=\int d^3\mathbf{x}'\frac{T^{00}(t-|\mathbf{x}-\mathbf{x}'|)}{|\mathbf{x}-\mathbf{x}'|} \\
\label{eq:AJ}
A^j(t,\mathbf{x})&=\int d^3\mathbf{x}'\frac{T^{0j}(t-|\mathbf{x}-\mathbf{x}'|)}{|\mathbf{x}-\mathbf{x}'|} 
\end{align}
For slow-motion objects, the metric \eqref{eq:metric} is sufficient to describe gravitational force up to  $v/c$ order, given by 
\begin{equation}
\mathbf{F}_g = M (\mathbf{E}+4\mathbf{v}\times \mathbf{B})
\end{equation}
with 
\begin{align}
\mathbf{E}& =-\nabla\Phi -\partial_t \mathbf{A}\,, \\
\mathbf{B}&=\nabla\times \mathbf{A}
\end{align}
This is the standard treatment of linearized gravity, which is
Lorentz invariant and causal. Note that at this order, we only need to
explicitly address {\it energy conservation}, while the requirement of {\it momentum conservation} is only required at the next order. 

For objects that stay within the near zone of each other, we can use the gravitational version of the Coulomb and Biot-Savar Laws, given by 
\begin{align}
\Phi(t,\mathbf{x})&=\int d^3\mathbf{x}'\frac{T^{00}(t,\mathbf{x}')}{|\mathbf{x}-\mathbf{x}'|} \\    
A^j(t,\mathbf{x})&=\int d^3\mathbf{x}'\frac{T^{0j}(t,\mathbf{x}')}{|\mathbf{x}-\mathbf{x}'|}
\end{align}
In this approximation, the effect of gravity is {\it instant}.  If we would like to
obtain the phase delay between the motion of the source and the
force exerted on a nearby object, we will have to revert to
Eqs.~\eqref{eq:PhiA} and \eqref{eq:AJ}.  Expanding those
equations to $(v/c)^5$ will lead to the first non-vanishing phase
delay at 2.5 post-Newtonian order.  This will lead to the remarkable
result for two objects separated by distance $L$: when one object
has its position oscillate at frequency $\omega$, the other object
will feel a force that oscillates at the same frequency $\omega$, but with a phase delay of $(\omega L/c)^5$. This makes the propagation effect of gravity very challenging to observe in the lab. 

\subsection{Energy Conservation for a Quantum Mass}
\label{sec:quantum_case}

In the quantum case, we define the stress-energy tensor operators
for a point particle with position operator $\hat X$ and momentum
operator $\hat P$ as
\begin{align}
    \label{eq:T00op}
    \hat T^{00}(t,x) &= M \delta(x-\hat X)\,, \\
    \label{eq:T01op}
    \hat T^{01}(t,x) &= \frac{1}{2} \left( \hat P \delta(x-\hat X) + \delta(x-\hat X) \hat P \right)\,.
\end{align}
Taking expectation values on the wavefunction $\psi(x,t)$, we
obtain the quantum stress-energy tensor components
\begin{align}
    \label{eq:T00exp}
    \langle \hat T^{00}(t,x) \rangle &= M |\psi(x,t)|^2\,, \\
        \label{eq:T01exp}
    \langle \hat T^{01}(t,x) \rangle &= \hbar \operatorname{Im}\left(\psi^*(x,t) \frac{\partial \psi(x,t)}{\partial x}\right)\,.
\end{align}
We can see that  $\partial_\nu \langle \hat T^{0\nu} \rangle$
vanishes as long as $\psi(x,t)$ satisfies a Schr\"odinger equation with a Hamiltonian whose kinetic part is $\hat P^2/(2M)$ with a potential that depends only on $x$ and $t$.  This corresponds well to Eq.~\eqref{eq:class:energy_conservation} in the classical case, where energy~\footnote{This is the total energy that includes the rest mass.} conservation $\partial_\nu T^{0\nu} = 0$ is satisfied as long as the momentum of the mass is equal to $m \dot X$.

Let us now turn to a macroscopic object with mass $M$ and a matter distribution profile $\rho(x)$, with $x$ the relative coordinate with respect to the center of mass (CM).  Suppose the CM position and momentum operators are $\hat X$ and $\hat P$, respectively, then the stress-energy tensor operators are given by
\begin{align}
    \hat T^{00}(t,x) &= \rho(x-\hat X)\,, \\
    \hat T^{01}(t,x) &= \frac{1}{2M} \left( \hat P \rho(x-\hat X) + \rho(x-\hat X) \hat P \right)\,.
\end{align}
Here we have adopted a treatment in Ref.~\cite{yang2013macroscopic}, where $\rho$ is the expectation value over the internal quantum state of the atoms and electrons that make up the object, and is characterized by narrow peaks around the lattice sites.  Note that in $T^{01}$ we have adopted Weyl quantization, which is the symmetrized version of the classical expression.  

In terms of the CM wavefunction $\psi(x,t)$, we have
\begin{align}
    \langle \hat T^{00}(t,x) \rangle & = \int dx' \rho(x-x') |\psi(x',t)|^2\,, \\   
    \langle \hat T^{01}(t,x) \rangle & = \int dx' \rho(x-x') \frac{\hbar}{M} \operatorname{Im}\left(\psi^*(x',t) \frac{\partial \psi(x',t)}{\partial x'}\right)\,.
\end{align}
Once more, the energy conservation condition is satisfied as long as the CM wavefunction satisfies a Schr\"odinger equation with a Hamiltonian whose kinetic part is $\hat P^2/(2M)$ with a potential that depends only on $x$ and $t$.

In reality, we almost never  deal with an object in a conservative potential well; objects instead interact with other objects, and EM fields that are used to manipulate and measure them.  We can in principle write down a multi-dimensional Schr\"odinger equation for the joint wavefunction of all the objects, and the energy conservation condition will be satisfied as long as each object has a kinetic term of the form $\hat P^2/(2M)$ and the potential depends only on the positions and time.  However, a more convenient way is to go to the Heisenberg picture, where the operators evolve according to the Heisenberg equation of motion, and the state is fixed at the initial time.  We shall do this in the next section, which also allows us to incorporate momentum conservation in a straightforward way.

\subsection{Energy-Momentum Conservation in the Heisenberg Picture}
\label{sec:quantum_case:momentum}

Even though momentum conservation is not required at the leading $v/c$ order, it is still interesting to see how it is satisfied in the quantum case. For a point particle in 1-D,  we will understand Eqs.~\eqref{eq:T00op} and \eqref{eq:T01op} as Heisenberg operators, plus a Weyl-symmetrized version of the $T^{11}$ operator, given by
\begin{align}
    \hat T^{11}(t,x) &= \frac{\hat P^2 \rho(x-\hat X) + 2\hat P \rho(x-\hat X) \hat P + \rho(x-\hat X) \hat P^2}{4M^2} \nonumber\\
    & +\frac{\hat F \Xi(\hat X-x)+ \Xi(\hat X-x )\hat F}{2M}
\end{align}
where $\hat F$ is the Heisenberg operator of the force acting on the particle, and  $\Xi'(x) = \rho(x)$ corresponds to how the force $\hat F$ is distributed onto different parts of the mass.~\footnote{The current way distributes the force evenly at each unit mass, and therefore causes no internal stress within the mass.  More realistically, force is exerted near the surface of the object, which causes internal deformations and excites internal oscillations.  We have adopted the simplified model here so we do not have to treat those internal physics of the mass.}

In the Heisenberg picture, since the quantum state does not evolve, energy momentum coservation simply requires the divergence-free of the $T^{\mu\nu}$ operator. Explicitly, for energy conservation, we have
\begin{equation}
    \frac{\partial}{\partial t} \hat T^{00} = \frac{1}{2M} \left( \hat P \rho'(x-\hat X) + \rho'(x-\hat X) \hat P \right) = -\partial_x \hat T^{01}
\end{equation}
which is satisfied as long as the kinetic term of the Hamiltonian is $\hat P^2/(2M)$ and the potential term depends only on $x$ and $t$.  Note here that the partial $t$ also takes times derivatives of Heisenberg operators.  We have used  $d\hat O/dt = i[\hat H, \hat O]$, the commutation relation $[\hat X, \hat P] = i\hbar$,  the fact that $[\hat P^2,\rho(x-\hat X)]=\hat P [\hat P,\rho(x-\hat X)]+[\hat P,\rho(x-\hat X)] \hat P$, and the fact that $[\hat P,f(\hat X)]=i\hbar f'(\hat X)$.

For momentum conservation, we can similarly verify that    
\begin{align}
    \frac{\partial}{\partial t} \hat T^{01} & = \frac{1}{4M^2} \left( \hat P^2 \rho'(x-\hat X) + 2\hat P \rho'(x-\hat X) \hat P + \rho'(x-\hat X) \hat P^2 \right) \nonumber\\
    & +\frac{\hat F \rho(x-\hat X)}{M} = -\partial_x \hat T^{11}
\end{align}
Here we have broken $\hat H$ into 
\begin{equation}
    \hat H = \frac{\hat P^2}{2M} + V(\hat X,t) - \hat F \hat X
\end{equation}
and have assumed $\hat F$ commutes with $\hat P$ and $\hat X$.

\subsection{Summary}

In this section, after discussing the energy-momentum budget of a classical test mass being moved by a light beam, we focused on linearized gravity that mediates interaction between slowly moving sources.  Up to $O(v/c)$ order, the gravitational field can be consistently sourced by the expectation value of the stress-energy tensor on the wavefunction as long as energy conservation is satisfied, which is guaranteed by the Schr\"odinger equation.

However, when the system is subject to continuous monitoring,
$\psi$ no longer satisfies the free Schr\"odinger equation.  In
Kent's causal-conditional formulation (Sec.~II), the local pure
state $|\psi(x)\rangle$ acquires a nontrivial spacetime
dependence through the measurement record, and the resulting
conditional expectation value
$\langle \psi(x)|\hat{T}^{\mu\nu}|\psi(x)\rangle$ is no longer
divergence-free.  Resolving this energy-momentum non-conservation,
while preserving the causal structure already secured, is the
central task of the next section.

\section{Energy-momentum Non-Conservation and the Corrector Field}
\label{sec:corrector}

In this section, we first diagnose the
physical origin of the non-conservation of the expectation value of the stress-energy tensor, and then introduce a
corrector field that restores $\nabla_\nu T^{\mu\nu} = 0$ while
producing only very weak post-Newtonian corrections to the gravitational
field.

\subsection{Why is Energy-Momentum Not Conserved}

In the CCSN framework, the gravitational field is sourced by the
conditional expectation value of the stress-energy tensor on the
local pure state.  Recalling the linearized gravity formulation of
Sec.~\ref{sec:classical_em}, in the Lorenz gauge we have
\begin{equation}
\square \bar h^{\mu\nu}(x^\rho) = -16\pi G \langle \psi(x^\rho) | \hat{T}^{\mu\nu}(x^\rho) | \psi(x^\rho) \rangle
\end{equation}
The crucial difference from the standard Schr\"odinger-Newton theory
of Sec.~\ref{sec:quantum_case} is that the local pure state
$|\psi(x^\rho)\rangle$ now depends on the spacetime position $x^\rho$ through
the measurement record in the past light cone.  This
spacetime dependence means that
\begin{equation}
    T_{\rm SN}^{\mu\nu}(x^\rho) \equiv \langle \psi(x^\rho) | \hat{T}^{\mu\nu}(x^\rho) | \psi(x^\rho) \rangle
\end{equation}
 is not
guaranteed to have a vanishing divergence, even though the
underlying operator $\hat{T}^{\mu\nu}(x^\rho)$ is conserved at the quantum
level.  Since the Einstein equation requires a divergence-free
source, we must supplement the conditional stress-energy tensor with
a corrector field.  Specifically, we seek a total stress-energy
tensor 
\begin{equation}
    \label{eq:total_T}      
T^{\mu\nu}_{\rm tot}(x^\rho) =T_{\rm SN}^{\mu\nu}(x^\rho) + T_{\rm corr}^{\mu\nu}(x^\rho)
\end{equation}
such that $\nabla_\nu T_{\rm tot}^{\mu\nu}(x^\rho) = 0$.
We will show that such a corrector field can be constructed using the standard procedure of Stuckelberg fields, and that it produces only very weak post-Newtonian corrections to the gravitational field.  We will only explore the consequences for slowly moving sources, for which we will only need to enforce
\begin{equation}
\partial_\nu T^{0\nu}_{\rm tot} = 0\,.
\end{equation}

\subsection{Physical Origin of the Non-Conservation and its different Interpretations}

Let us first understand the physical origin of the non-conservation of $T_{\rm SN}^{\mu\nu}$.  In particular, we recall that its conservation is not required with standard quantum mechanics --- even though the ``meaning'' of this non-conservation differs by interpretation. 

As we recall from Sec.~\ref{sec:quantum_case}, mass-energy conservation is satsified whenever $\psi(x,t)$ evolves according to the Schr\"odinger equation --- yet this is {\it broken} when we perform a measurement:
\begin{equation}
    \frac{d\langle X(t)\rangle}{dt}  \neq \frac{\langle P(t) \rangle}{M} 
\end{equation}

In this section, we shall start from energy non-conservation in a single measurement, and then move on to the continuous monitoring case. 

\subsubsection{A Single Measurement} 
\label{sec:single_measurement}

Let us consider a  test particle with initial wavefunction $\psi_0(x)$ centered around $x=0$ with a width $\sigma_0$, with $\hat x$ imprecisely measured with an error of $\sigma_M \ll \sigma_0$.  This is realized by coupling the mass with an ancilla system (for example an optical field), and then performing a projective measurement on the ancilla.  More specifically, if $(\hat x, \hat p)$ and $(\hat Q,\hat P)$ are the position and momentum operators of the test particle and the ancilla, respective, and the unitary evolution operator for the coupling is given by $U =\exp(-i  \hat x \hat P/\hbar)$, then given an initial joint state of 
\begin{equation}
|\Psi_{\rm init}\rangle = \int dx dQ    \psi_0(x) \phi_0(Q) |x\rangle\otimes |Q\rangle
\end{equation}
the final joint state will become
\begin{align}
    \label{eq:finalstate}
|\Psi_{\rm final}\rangle &= \int dx dQ    \psi_0(x) \phi_0(Q- x) |x\rangle\otimes |Q\rangle\nonumber\\
& = \int dx dQ    \psi_0(x) \phi_0(Q) |x\rangle\otimes |Q+ x\rangle \nonumber\\
&=  \int d\xi dQ     \psi_0(Q-\xi) \phi_0(\xi) |Q-\xi \rangle\otimes |Q\rangle
\end{align}
From the second line of the above eqution, we see that the ancilla's position $Q$ follows the sum distribution associated with $\psi_0$ and $\phi_0$.  From the third line, we see that upon detection $\tilde Q$ for the ancilla, the test particle location $x'$ is centered approximately around $Q$ with a width of $\approx \sigma_M$.  The sudden jump of the position of the mass is not balanced by the mass ever having a large momentum, and therefore ``defies'' energy conservation.  Even though the measuring device will impart momentum back onto the mass, that is {\it not} the origin of this non-conservation.  For example, the non-conservation of energy is more serious when the initial width of the wavefunction $\sigma_0$ is larger, while the momentum kick from the measuring device is unrelated to $\sigma_0$.   In fact, momentum kick from the device is the direct reason why the {\it width} of the wavefunction shrinks to $\sim \sigma_M$ which corresponds to a larger momentum uncertainty.  

In the {\it Copenhagen Interpretation}, measurement is a non-unitary process and not governed by the Schr\"odinger equation. Energy conservation is not supposed to be enforced during the measurement process, and therefore the non-conservation of energy does not pose inconsistency within the predictable power of the theory. 

In the {\it Many World Interpretation}, when the ancilla is projectively measured, the system and ancilla become entangled further with the observer and the environment, with the wavefunction of the universe given as a sum over all possible results of measurement, each corroborated by a particular state of the observer and environment: 
\begin{align}
    \label{eq:manyworld}
|\Psi_{\rm MW}\rangle &=  \int d\xi dQ     \psi_0(Q-\xi) \phi_0(\xi) |Q-\xi \rangle\otimes |Q\rangle \otimes |Q\rangle_{\rm obs\,env}\,.
\end{align}
Conservation of energy and momentum is satisfied by $|\Psi_{\rm MW}\rangle$ as a whole, but not by any particular branch that the observers find themselves in.  However, since quantum mechanics does not predict which branch the observers will find themselves in, the non-conservation of energy and momentum in each branch does not pose inconsistency within the predictable power of the theory.

\subsubsection{Continuous Measurement}

Consider the simplest continuous measurement process in
which in-going light field quadratures $\hat a_{1,2}$ are incident on a
test mass, measuring it with strength $\alpha$.  This process is widely treated, and we will use the notation of Ref.~\cite{chen2013macroscopic}.  In the Heisenberg
picture, the Heisenberg operators of the out-going field
quadratures $\hat b_{1,2}$ are given by
\begin{equation}
    \hat b_1=\hat  a_1\,,\quad \hat  b_2 = \hat  a_2 + \alpha \hat  X\,,
\end{equation}
while those of the test mass's position and momentum are given by
\begin{equation}
    \dot {\hat X} = \hat P/M\,,\quad \dot {\hat P} = \alpha \hat a_1\,,
\end{equation}
with $\alpha \hat a_1$ the radiation pressure force, which enforces measurement back action. This set of equations is the quantum mechanical counterpart of the classical equations of motion. Since the rate of change of the position of the mass is determined by its momentum, and the change of the momentum of the mass is balanced by the stress of the
field, $\hat T^{\mu\nu}$ at the quantum level is conserved.  

We take the measurement record to be obtained by homodyne detection of the outgoing quadrature $\hat b_2$, which carries the position information.
A continuous measurement of $\hat b_2$ will cause the state of the system to undergo a random walk, and therefore the expectation value of $\hat T^{\mu\nu}$ on the post-measurement state will not be conserved.  The Stochastic Schr\"odinger equation for the state of the system is given by
\be
|d\psi\rangle = -\frac{i}{\hbar} \hat{H}_0|\psi\rangle dt + \frac{\alpha (\hat{X} - X_c(t))}{\sqrt{2}} |\psi\rangle dW
- \frac{\alpha^2}{4} (\hat{X} - X_c(t))^2 |\psi\rangle dt\,.
\ee
where $|\psi\rangle$ is the {\it conditional quantum state} of the system, which depends on the measurement record, and $dW$ is a Wiener increment.  The conditional expectation value of the position and momentum operators are given by
\begin{equation}
X_c(t) = \langle \psi | \hat X | \psi \rangle\,,\quad P_c(t) = \langle \psi | \hat P | \psi \rangle\,.
\end{equation}
They evolve in time according to 
\begin{equation}
    \label{eq:stochastic_evolution}
    \dot {X} _c = P_c/M + \sqrt{2}{\alpha V^c_{xx}} dW/dt\,,\quad \dot {P}_c = \sqrt{2}\alpha V^c_{xp} dW/dt
\end{equation}
where $V^c_{xx}$ and $V^c_{xp}$ are the conditional variance of the position and covariance between position and momentum, respectively, evaluated on the conditional state $|\psi\rangle$.

For a general mass with matter density profile $\rho(x)$ (reduces to $\delta(x)$ for a point particle), inserting the stochastic Schr\"odinger equation
\be
|d\psi\rangle = -\frac{i}{\hbar} \hat{H}_0|\psi\rangle dt + \frac{\alpha (\hat{X} - X_c(t))}{\sqrt{2}} |\psi\rangle dW
- \frac{\alpha^2}{4} (\hat{X} - X_c(t))^2 |\psi\rangle dt
\ee
into Eqs.~\eqref{eq:T00exp} and \eqref{eq:T01exp}, we can see that $\partial_\nu\langle T^{0\nu}\rangle$ is non-zero, even when $\hat{H}_0$ is the Hamiltonian of a free mass. In particular, for $\langle\hat{T}^{00}\rangle$, fixing $x$ and incrementing $t$ by $dt$, we have
\be
d \langle \hat{T}^{00} \rangle =  \langle d\psi| \rho(x-\hat X) |\psi\rangle + \langle \psi | \rho(x-\hat X) |d\psi\rangle + \langle d\psi| \rho(x-\hat X) |d\psi\rangle\ee 
where the last second-order term is non-negligible due to the Ito calculus $dW^2 = dt$. 
The $dt$-terms generated by the measurement update cancel explicitly:
\begin{align}
    &-\frac{\alpha^2}{4}\left\langle\left\{(\hat X-X_c)^2,\rho(x-\hat X)\right\}\right\rangle dt\nonumber\\
    &\quad+\frac{\alpha^2}{2}\left\langle(\hat X-X_c)\rho(x-\hat X)(\hat X-X_c)\right\rangle dt
    =0,
\end{align}
because $\rho(x-\hat X)$ commutes with $\hat X-X_c$. Thus only the stochastic contribution proportional to $dW$ remains. The measurement-induced part of the increment is
\begin{align}
    d\langle \hat T^{00} \rangle
    =\sqrt{2}\alpha \left\langle(\hat X-X_c)\rho(x-\hat X)\right\rangle dW.
\end{align}
The unitary part cancels against $\partial_x\langle \hat T^{01}\rangle$, as in the unmeasured case, yielding
\be
\begin{split}
\partial_\nu\langle T^{0\nu}\rangle=
\sqrt{2}\alpha \langle \psi| (\hat{X} - X_c(t)) \rho(x-\hat X)|\psi \rangle dW/dt
\end{split}
\ee

\subsubsection{Narrow Wavefunction Expansion}

In the situation where  $\Delta x_{\rm CM}$ of the CM wavefunction is small compared with the scale at which $\rho$ varies --- meaning, the CM moves less than the zero-point motion of nuclei  $x_{\rm zp}$  around the lattice sites --- we can expand
\begin{equation}
    \label{eq:rho_expansion}
    \rho(x-\hat X) = \rho(x-X_c(t))+ \rho'(x-X_c(t))(\hat X - X_c(t))+ \ldots.
\end{equation}
We shall refer to this as the ``narrow-wavefunction'' expansion; subsequent terms will contain higher spatial derivatives of $\rho$, and are suppressed by higher powers of $\Delta x_{\rm CM}/x_{\rm zp}$.
For a harmonically trapped CM wavefunction with trap frequency $\Omega_q$, $\Delta x\sim\sqrt{\hbar/(M\Omega_q)}$. Using $x_{\rm zp}\sim\sqrt{\hbar/(m\omega_{\rm ph})}$, where $m$ is the nuclear mass and $\omega_{\rm ph}$ is a characteristic phonon frequency, the narrow-wavefunction condition becomes $M\Omega_q\gg m\omega_{\rm ph}$.

Using this expansion, we can obtain formulas for energy and momentum conservation that can be interpreted more easily, by directly expanding $\langle T^{\mu\nu}\rangle$. Taking the leading two terms from the expansion~\eqref{eq:rho_expansion}, we can write
\begin{align}    
    \langle \hat T^{00}(t,x) \rangle &\approx \rho(x-X_c(t))\,,\; \\
        \langle \hat T^{01}(t,x) \rangle &\approx \frac{P_c(t) \rho(x-X_c(t))}{M}+\frac{V_{xp}^c (t)}{M}\rho'(x-X_c(t))\\
    \langle \hat T^{11}(t,x) \rangle & \approx \frac{\langle\psi |\hat P^2| \psi \rangle \rho(x-X_c(t))}{M^2} + \frac{F_c(t) \Xi(X_c -x)}{M }\,,
\end{align}
where $P_c(t) = \langle \psi | \hat P | \psi \rangle$ is the conditional expectation value of the CM momentum, $V_{xp}^c(t)$ the conditional covariance between position and momentum, and $F_c(t) = \langle \psi | \hat F | \psi \rangle$ the conditional expectation value of the force acting on the mass.  Note also that $\rho$ is one order higher in the narrow wavefunction expansion than $\Xi$ since $\rho = \Xi'$.

Up to leading order in narrow-wavefunction expansion, we obtain 
\begin{align}
    \label{eq:nonconservation_narrow}
    \partial_\nu \langle T^{0\nu} \rangle &= -\left( \dot{X}_c(t) - \frac{P_c(t)}{M} \right) \rho'(x-X_c(t))\,,
    \\
        \partial_\nu \langle T^{1\nu} \rangle& = (\dot P_c- F_c) \frac{\rho(x-X_c(t))}{M}   \,,
\end{align}   
We can clearly see that the non-conservation of energy is associated with the mismatch between the rate of change of the expectation value of the position of the mass and the expectation value of its momentum, while the non-conservation of momentum is associated with the mismatch between the rate of change of the expectation value of the momentum and the expectation value of the force acting on the mass.

The two non-conservation equations have closely related but distinct meanings. The quantity $\dot X_c-P_c/M$ measures an anomalous velocity of the conditional mass distribution: the center of the conditional
density moves at a rate not accounted for by the conditional momentum
current. Similarly, $\dot P_c/M-F_c/M$ may be viewed as an anomalous
momentum acceleration, arising because the measurement record updates
the conditional expectation value of momentum in addition to mechanical force. This should be distinguished from the anomalous
acceleration of the conditional trajectory, $\ddot X_c-F_c/M$, which
combines the time derivative of the anomalous velocity with the
anomalous momentum acceleration. If we denote the anomalous velocity
$w=-(\dot X_c-P_c/M)$ and momentum-level anomalous acceleration $b=\dot P_c/M-F_c/M$, this trajectory-level
anomalous acceleration is therefore $b-\dot w$.

Inserting evolution equations for $X_c$ and $P_c$, or Eqs.~\eqref{eq:stochastic_evolution}, we obtain 
\begin{align}
    \label{eq:nonconservation_narrow_explicitx}
    \partial_\nu \langle T^{0\nu} \rangle &= -\sqrt{2}\alpha V_{xx}^c(t) \rho'(x-X_c(t )) dW/dt\,,\\
        \label{eq:nonconservation_narrow_explicitp}
    \partial_\nu \langle T^{1\nu} \rangle &= \sqrt{2}\alpha \frac{V_{xp}^c(t)}{M} \rho(x-X_c(t)) dW/dt\,.
\end{align} 
This is an additional $v/c$ suppressed from the non-conservation of energy, but lacks a suppression from the width of the wavefunction over the zero point fluctuation of nuclei around the lattice sites. 

From this simple 1+1 dimensional model, we can summarize  the non-conservation of $T^{\mu\nu}$ in 3+1 dimensional case as 
\begin{equation}
\nabla_\nu T^{0\nu}_{\rm SN} =\frac{w^i (t)}{c} \partial_i \rho(\mathbf{x} - \mathbf{X}_c(t)) =\frac{1}{c}\mathbf{w}(t) \cdot \nabla \rho(\mathbf{x} - \mathbf{X}_c(t))\,,
\end{equation}
where $w^i (t)$ is the 3-dimensional anomalous velocity. Note that if the measurement is only performed along the $x$ direction, then the vector $\mathbf{w}$ will only have an $x$ component.

Note that Eq.~\eqref{eq:nonconservation_narrow_explicitx} has a very suggestive form: the non-conservation of energy is associated with the mismatch between the rate of change of the expectation value of the position of the mass and the expectation value of its momentum.  This difference is because the rate of change of the position also arises from the gain of information from the measurement, which does not arise from the expectation value of the momentum --- like we had discussed in Sec.~\ref{sec:single_measurement}. Indeed,  this difference is larger if we measure the position more strongly (larger $\alpha$) and if the wavefunction is more spread out (larger $V^c_{xx}$).  The latter point is consistent with the intuition from the single measurement case, where the non-conservation of energy is more severe when the initial wavefunction is wider.  Compared with the single measurement case, continuous monitoring causes a continuous non-conservation of energy along the world line of each mass.  We do note that $dW/dt$ is formally infinite, but this is because we have assumed measurement to take place at all time scales; appropriately cutting if off at the bandwidth of the measurement will suppress the high-frequency fluctuations in energy non-conservation.

At steady state, if we band-limit the white noise $dW/dt$ at a measurement bandwidth of  $\Lambda$, which leads to an rms value of $\sqrt{\Lambda}$ for $dW/dt$, we will have
\begin{equation}
|\mathbf{w}|\sim \sqrt{\frac{\hbar \Lambda}{M}}\,.
\end{equation}
Even though $dW/dt$ is formally infinite, it is very hard for $\mathbf{w}$ to become a sizable fraction of the speed of light in actual measurement scenarios.  For example, for a nanoparticle with density $3\times 10^3\,{\rm kg}/{\rm m}^3$ with radius 10\,nm, measured with a bandwidth of $\Lambda = 10^{12}\,{\rm Hz}$, we have $w/c \sim 10^{-9}$, still very small. 

For momentum non-conservation, we can write
\begin{equation}\nabla_\nu T^{k\nu}_{\rm SN} =\frac{\mathbf{b}(t)}{c^2}\,\rho(\mathbf{x} - \mathbf{X}_c(t))\,,
\end{equation}
with $\mathbf{b}$ having the dimension of acceleration, and a typical scale of
\begin{equation}\mathbf{|b|} \sim \Omega_q \sqrt{\frac{\hbar \Lambda}{M}}\,.
\end{equation}

\subsubsection{Structure of the light-mass wavefunction before collapse}
The same physics can be explained without state collapse, if we consider the joint Hilbert space of the mass and out-going optical field. 
\begin{widetext}
Up till any time $t$, we have a joint quantum state, between three systems: the mass, the out-going light so far, and the in-going light that is yet to interact with the mass.  This state vector can be written as a path integral over all measurement histories $\{\xi(t'):t'< t\}$: 
\begin{equation}
|\Psi(t)\rangle =  \int_{-\infty}^t D[\xi]  \underbrace{ \big|\psi_c[\xi(t')]\big\rangle_M }_{\rm mass }\otimes \underbrace{\big|0:\; t'>t\big\rangle_{\rm in}}_{\rm ingoing\,light}\otimes \underbrace{\big| b_2(t') = \xi(t'): \;t'<t\big\rangle_{\rm out} }_{{\rm outgoing\, light}}\,.
\end{equation}
\end{widetext}
Here, the ``in-going'' part of the light field is yet to interact with the mass, hence enters the wavefunction of the light-mass system as a product. Here we have assumed that this part of light is at vacuum, but it could have been prepared into other states.  
The mass and the out-going field have interacted and hence is no longer at a product state.  This part of the state can be written in different ways; we have chosen to express it as a sum over simultaneous eigenstates of the quadrature operators $\hat b_2(t')$ with $t'<t$, in which case the associated  state of the mass is the {\it conditional state} given the measurement history $\xi(t')$.   This sum is over all  possible measurement histories $\xi(t')$ up to time $t$.  As time goes on, more light degress of freedom will have interacted with the mass; they get moved out of the ``ingoing light'' part and goes into the ``outgoing light'' part. 

The non-conservation of energy-momentum is associated with  the conditional state of the mass, but not by the state of the universe that contains all the possible outcomes.  Even though we only obtain one particular experimental outcome each time we perform the experiment, all the other branches of the universe's wavefunction is still a conceptual necessity if we would like to maintain energy-momentum conservation.  

More generally, as a general set of measurements were performed on out-going operators $\hat Z(t)$, the joint light-mass quantum state $|\Psi(t)\rangle$ remains unchanged, but will be decomposed into a sum of different terms $|\hat Z(t') = \zeta(t')\rangle_{\rm out}$, corresponding to simultaneous eigenstates of the new operators being measured. What remains unchanged is the density matrix $\hat\rho_M(t)$ of the mass if we trace out the out-going fields.  Density matrix $\hat\rho_M(t)$ satisfies the same master equation; different measurement processes correspond to Stochastic Schr\"odinger Equations that are different {\it stochastic unraveling} of the same master equation.  

\subsection{Corrector Field}

The prominent feature of CCSN theory is that  the particular way the joint light-mass wavefunction $|\Psi(t)\rangle$ is broken into branches --- due to the measurement of $\hat Z(t)$ --- affect the gravitational field, even though each branch does not conserve energy-momentum. However, allowing all the other branches to contribute to energy-momentum will cause an Everett Telephone effect highlighted by Page and Geilker~\cite{page1981indirect}. To resolve this dilemma, let us construct a corrector field that 
compensates for the non-conservation of energy-momentum. Our goal is to write
\begin{equation}
G^{\mu\nu} = 8\pi G \left( \langle \psi(x) | \hat{T}^{\mu\nu}_{\rm SN} | \psi(x) \rangle + T_{\text{corr}}^{\mu\nu} \right)
\end{equation}
in such a way that $T^{\mu\nu}_{\text{corr}}$ has a non-zero
divergence that exactly cancels the non-zero divergence of
$\langle \psi(x) | \hat{T}^{\mu\nu}_{\rm SN} | \psi(x) \rangle$.
In this way, we can restore energy-momentum conservation, in addition to
causality, in semiclassical gravity.  

Our corrector field formulation was motivated by the
St\"ukelberg mechanism used in quantum electrodynamics with
massive photons, where a scalar field is introduced to restore
gauge invariance \cite{stuckelberg1938,ruegg2004}.  In our case, we will introduce a vector field
to restore energy-momentum conservation.
The corrector field should not be interpreted as a new physical matter field with independently specifiable degrees of freedom.  Its role is analogous to that of a St\"uckelberg field in the sense that an auxiliary field is introduced to restore a desired structural property of the equations. Once the non-conservation current $J^\mu$ of a conditional branch is given, $A^\mu$ is fixed by the field equation together with the chosen boundary condition.
Suppose the
non-conservation of
$T^{\mu\nu}_{\rm SN} = \langle \psi(x) | \hat{T}^{\mu\nu}_{\rm SN}
| \psi(x) \rangle$ is given by
\begin{equation}
\nabla_\nu T^{\mu\nu}_{\rm SN} = J^\mu
\end{equation}
where $J^\mu$ is a non-zero vector field.  We want to introduce a
corrector field, which is a vector field $A_\mu$, with 
\begin{equation}
\square A^\nu = -J^\nu
\end{equation}
We take $A^{\mu}$ to be the retarded solution of this equation and set the homogeneous solution to zero so that the corrector is fully determined by the past history of $J^\mu$.
We also introduce
\begin{equation}
C^{\mu\nu} = \nabla^\mu A^\nu + \nabla^\nu A^\mu - g^{\mu\nu} \nabla_\alpha A^\alpha
\end{equation}
so that
\begin{equation}
\nabla_\nu C^{\mu\nu} = -J^\mu
\end{equation}
We will then use $C^{\mu\nu}$ as the corrector field, and write
\begin{equation}
G^{\mu\nu} = 8\pi G \left( T^{\mu\nu}_{\rm SN} + C^{\mu\nu} \right)
\end{equation}
Let us check the Newtonian limit of this theory.  In the
Newtonian limit, we have
\begin{equation}
G^{00} = 8\pi G (T^{00}_{\rm SN} + C^{00})
\end{equation}
with 
\begin{equation}
    C^{00} =C_{00}= - \partial_0 A^0  +\partial_j A^j  \,.
\end{equation}  
a correction to matter density. 

\subsection{Corrector Field for Continuous Measurement and Narrow-Wavefunction Expansion}

Continue from the previous section with 
\begin{align}
J^0 = \nabla_\nu T_{\rm SN}^{0\nu} =\frac{w^i (t)}{c} \partial_i \rho(\mathbf{x} - \mathbf{X}_c(t))\,, \\
J^i = \nabla_\nu T_{\rm SN}^{i\nu} =\frac{b^i(t)}{c^2}\,\rho(\mathbf{x} - \mathbf{X}_c(t))\,,   
\end{align}
we can solve for $A^0$ and $A^i$, in the
slow-motion approximation, as 
\begin{align}
A^0(\mathbf{x},t)& = \int d^3 \mathbf{x}' \frac{J^0(\mathbf{x}',t)}{4\pi |\mathbf{x} - \mathbf{x}'|}\nonumber\\
&=\int d^3 \mathbf{x}' \frac{w^i (t)}{4\pi c} \frac{\partial_i \rho(\mathbf{x}' - \mathbf{X}_c(t))}{|\mathbf{x} - \mathbf{x}'|} \,,\\
A^i(\mathbf{x},t) & = \int d^3 \mathbf{x}' \frac{J^i(\mathbf{x}',t)}{4\pi |\mathbf{x} - \mathbf{x}'|} \nonumber\\
& = \int d^3 \mathbf{x}' \frac{b^i(t)}{4\pi c^2}\frac{\rho(\mathbf{x}' - \mathbf{X}_c(t))}{|\mathbf{x} - \mathbf{x}'|} \,.
\end{align}
This leads to 
\begin{align}
\partial_0 A^0 =
\int d^3 \mathbf{x}' \frac{\dot w^i (t)}{4\pi c^2} \frac{\partial_i \rho(\mathbf{x}' - \mathbf{X}_c(t))}{|\mathbf{x} - \mathbf{x}'|}\,,\\
\partial_j A^j = \int d^3 \mathbf{x}' \frac{b^j(t)}{4\pi c^2}\frac{\partial_i \rho(\mathbf{x}' - \mathbf{X}_c(t))}{|\mathbf{x} - \mathbf{x}'|} \,.
\end{align}
It is interesting that 
\begin{equation}
    \dot w^i -b^i =-\left(\frac{d^2 X_c^i(t)}{dt^2} - \frac{F_c^i(t)}{M}\right)\equiv- a_{\rm col}^i(t)\,,
\end{equation}
with $a_{\rm col}^i(t)$ the compensation for the previously mentioned trajectory-level anomalous acceleration of the mass, which is contributed purely by the collapse processes --- unaccounted for by the expectation value of the force.  This is not surprising since the non-conservation of energy and momentum is associated with the collapse process, and therefore the corrector field should be associated with the collapse process as well. We have
\begin{align}
    C_{00} (\mathbf{x},t)&= - \partial_0 A^0  +\partial_j A^j  = \frac{a_{\rm col}^i(t)}{4\pi c^2} \int d^3 \mathbf{x}' \frac{\partial_i \rho(\mathbf{x}' - \mathbf{X}_c(t))}{|\mathbf{x} - \mathbf{x}'|}
\end{align}
It is interesting to see that the integral over matter density yields a distribution that has the same behavior as the force exerted by the matter distribution $\rho$. This means $C_{00}$ decays at $1/r^2$ at large distances, leading to a finite total mass correction, and that it is finite at small distances.  

For the effect of $C_{00}$ on the mass, we will need to evaluate the potential energy of the mass in the gravity field generated by $C_{00}$, leading to $\hat V_{\rm corr}$ that we can insert into the SSNE in Sec.~\ref{subsec:SNNE}.  It is given by:
\begin{align}
\hat V_{\rm corr}(t) &=G \int d^3 \mathbf{x} d^3 \mathbf{y} \frac{\rho(\mathbf{x}-\hat{\mathbf{X}}) C_{00}(\mathbf{y})}{4\pi|\mathbf{x}-\mathbf{y}|}\nonumber\\
&=\frac{Ga_{\rm col}^i}{4\pi c^2} \int d^3 \mathbf{x}' d^3\mathbf{x} d^3\mathbf{y} \frac{\rho(\mathbf{x}-\hat{\mathbf{X}})\partial_i \rho(\mathbf{x}' - \mathbf{X}_c)}{|\mathbf{x} - \mathbf{y}||\mathbf{y} - \mathbf{x}'|} \nonumber\\
&=-\frac{G a_{\rm col}^i}{2c^2} \int d^3 \mathbf{x}  d^3\mathbf{y} |\mathbf{x}-\mathbf{y}| \rho(\mathbf{x}-\hat{\mathbf{X}}+\mathbf{X}_c)\partial_i \rho(\mathbf{y}) \nonumber\\
\end{align}
In the narrow-wavefunction expansion, we have 
which (skipping several steps) in the narrow wavefunction expansion can be written as
\begin{align}
V_{\rm corr}  \approx  I_{jk} \frac{a_{\rm col}^j (\hat X^k-X_c^k)}{2c^2}
\end{align}
with 
\begin{equation}
    I_{jk}=  G\int d^3\mathbf{x} d^3\mathbf{y} |\mathbf{x}-\mathbf{y}| \partial_k \rho(\mathbf{x})\partial_j \rho(\mathbf{y}) 
\end{equation}
If $\rho$ is the sum over $\mathcal{N}_{\rm site}$ lattice sites, each with a peak density of $\rho_{\rm zp}$ and zero point width $x_{\rm zp}$, then the above integral can be written as $\mathcal{N}_{\rm site}$ times the contribution from each lattice site, which is isotropic.  We will then have
\begin{equation}
    I_{jk} = -\frac{2\mathcal{N}_{\rm site} \delta_{jk}}{3} \int d^3\mathbf{x} d^3\mathbf{y} \frac{\rho_0(\mathbf{x})\rho_0(\mathbf{y})}{|\mathbf{x} - \mathbf{y}|}  \,.
\end{equation}
with $\rho_0$ the density profile near each lattice site. For a lattice site with nucleus mass $m$ distributed with a Gaussian profile with width $x_{\rm zp}$, we have 
\begin{equation}
    I_{jk} =-\frac{2G m^2 \mathcal{N}_{\rm site} }{3\sqrt{\pi} x_{\rm zp}} \delta_{jk}\,.
\end{equation}
Using the facts that
\begin{equation}
    \omega_{\rm SN}^2 = \frac{G m}{6\sqrt{\pi} x_{\rm zp}^3}\,,\quad M=m\mathcal{N}_{\rm site}\,,
\end{equation}
we obtain 
\begin{equation}
\hat V_{\rm corr} \approx -2 M \omega_{\rm SN}^2 x_{\rm zp}^2 \frac{a_{\rm col}^i (\hat X^i-X_c^i)}{c^2}
\end{equation}
This will be an {\it over estimate} of the effect of the corrector field, since interaction term is strongest when $\hat X$ is close to $X_c$, where the narrow wavefunction approximation is valid. 

This leads to a correction force proportional to 
\begin{equation}
    F_{\rm corr} \approx 2M \omega_{\rm SN}^2 x_{\rm zp}^2 \frac{a_{\rm col}}{c^2}
\end{equation}
Comparing this with 
\begin{equation}
    F_{\rm SN} \sim M\omega_{\rm SN}^2 \Delta x
\end{equation}
with $\Delta x \sim \sqrt{\hbar/(M\Omega_q)}$, the width of the wavefunction, and inserting the previous estimate
\begin{equation}
   | \mathbf{a}_{\rm col} |\sim |\mathbf{b}| \sim \Omega_q \sqrt{\frac{\hbar \Lambda}{M}}\,,
    \end{equation}
we have
\begin{equation}
    \frac{F_{\rm corr}}{F_{\rm SN}} \sim \frac{x_{\rm zp}^2 a_{\rm col}}{c^2  \Delta x}  \sim  \Omega_q^{3/2} \Lambda^{1/2} x_{\rm zp}^2/c^2
\end{equation}
This is very small. For example, for $\Omega_q \sim \Lambda = 2\pi\times 10^{12}$~Hz, and $x_{\rm zp} = 10^{-12}$~m, we have $\Delta F_{\rm corr}/F_{\rm SN} \sim 10^{-15}$. 

As an alternative way to estimate the effect of the corrector field, we can compute the acceleration induced by $F_{\rm corr}$, which is given by 
\begin{equation}
    \label{eq:acceleration_corrector}
    a_{\rm corr} \sim \frac{F_{\rm corr}}{M} \sim 2 \frac{\omega_{\rm SN}^2 x_{\rm zp}^2}{c^2} a_{\rm col}\,.
\end{equation}
With $\omega_{\rm SN} \sim 2\pi\times 0.1$~Hz, $x_{\rm zp} = 10^{-12}$~m, we have $\omega_{\rm SN}x_{\rm zp}/c \sim 10^{-21}$, and therefore $a_{\rm corr} \sim 10^{-42} a_{\rm col}$, which is completely negligible. We also note that this is likely an overestimate of the effect when the wavefunction is no longer narrow, since the effect of the force is the largest when the wavefunction is narrow.  We can define $v_{\rm SN} \equiv \omega_{\rm SN} x_{\rm zp}$ as the characteristic velocity of the Schr\"odinger-Newton potential (in fact its ``escape velocity'').  The fact that $v_{\rm SN}/c$ is very slow and ultra-non-relativistic is the reason why the corrector field's effect is negligible in current experiments.

Although the corrector effect is negligibly small, it is in principle experimentally testable within the same homodyne-measurement setup. The homodyne record of the outgoing quadrature $\hat b_2$ determines the conditional evolution of the test mass, so that $X_c(t)$ and $P_c(t)$, and hence the measurement-induced anomalous motion characterized by $a_{\rm col}(t)$, can be reconstructed causally from the corresponding filtering equations for a finite measurement bandwidth. The additional acceleration in Eq.~\eqref{eq:acceleration_corrector} is therefore not an arbitrary extra force, but a specific response tied to the same measurement record. In principle, one may propagate this predicted response through the mechanical and optical dynamics and analyze the measured homodyne record using models with and without the corrector field. Analogously to noise-weighted searches for weak signals, this allows one to define a signal-to-noise ratio, or more generally a likelihood ratio, between the two models.

However, Eq.~\eqref{eq:acceleration_corrector} shows that, within the regime considered here, the effect of the corrector field is much smaller than the collapse-induced effect itself, and therefore produces only a small correction even when the collapse-induced motion is appreciable. This means that previous predictions based on the CCSN formulation remain essentially unchanged, and that the corrector field is not expected to produce any observable effect in current experiments.

The discussion above focused on a continuously monitored self-gravitating test mass. The same corrector construction can also be applied to a localized measurement in the causally separated setup of Sec.~II B. In this case, the measurement on one subsystem changes the conditional stress-energy tensor of the distant subsystem only after the corresponding measurement record enters its past light cone. During this causal conditional-state update, the resulting stress-energy tensor is generally not locally conserved, giving rise to a nonzero $J^\mu$ and hence to a transient retarded corrector field. We discuss this complementary regime and its possible experimental signature in Appendix~B.

\section{Conclusion}
\label{sec:conclusion}

In this paper, we first acknowledged that a naive collapse of quantum states
in ``nonlinear quantum mechanics'' can lead to causality violation. By examining Kent's fix for restoring causality
in nonlinear quantum mechanics, we showed that the local pure state is simply a particular choice of Tomonaga-Schwinger hypersurface, and that this fix is mathematically equivalent to introducing a causal feedback control system. The remaining issue for mathematical consistency, however, is that the position dependence of the local pure state leads to a non-conservation of energy-momentum.

We then showed how to restore
energy-momentum conservation in semiclassical gravity by
introducing a corrector field.  The corrector field compensates
the ``missing momentum'' of a quantum trajectory.   We further showed  that in the slow-motion limit, the corrector field
only introduces very weak post-Newtonian corrections, and is therefore
consistent with current observations.  However, if measurements
are made at a large distance from the source, the corrector
field's effect can be larger, although to be observable, certain
coherence conditions must be met, and experiments that rule out
these effects can be extremely challenging. 

The restoration of energy-momentum conservation and the weakness
of the corrector field's effect allow us to consistently proceed
with experimental tests of semiclassical gravity, which has the
prominent testable signature of a self-gravitating wavefunction
and has effects several orders of magnitude larger than
gravity-induced entanglement. 

We emphasize that in the CCSN formulation, the measurement
process---{\it not axiomatically defined yet pragmatically
recognized in all experimental situations}---enters nontrivially
into the generation of the gravitational field. This is a sharp contrast with other classical theories of gravity, which separates the generation of gravity from the measurements the experimentalists are performing (for example in order to test the nature of gravity) and instead introduce additional observers to extract classical information and generate gravity independently~\cite{bahrami2014schrodinger,nimmrichter2015stochastic,tilloy2016sourcing}. 

However, unlike the original Schr\"odinger-Newton theory, which explicitly mentions the collapse of wavefunction across constant-time slices, CCSN formulates classical gravity as a feedback process sourced by measurement records, and hence fits into all interpretations of quantum mechanics. 
In particular, in the many-world interpretation, since CCSN only draws data from within each branch that has a unique measurement outcome, it does not allow communication across branches; this differs from Ref.~\cite{kaplan2022causal}.

Tomizuka and Takeda have recently argued and demonstrated that semiclassical gravity theories~\cite{oppenheim2023postquantum} can be fit into the framework of a larger quantum theory with decoherence~\cite{tomizuka2026emergence}.   We believe this argument is quite general and applies to not only to Refs.~~\cite{bahrami2014schrodinger,nimmrichter2015stochastic,tilloy2016sourcing} but also to CCSN theory. Ultimately, a unitary evolution within a Hilbert space plus a measurement process that satisfy the Born's rule are used in all these theories.  However, as we mentioned above, the distinction between CCSN and other semiclassical theories is the role played by quantum measurements. 
In this way, in CCSN, the nature of
quantum measurement (i.e., what is a measurement) remains at the heart of this problem.  For a deeper examination of this problem, one could resort to objective collapse models~\cite{bassi2023collapse} or Quantum Darwinism arguments~\cite{zurek2009quantum}.  By connecting testing the nature of gravity to exploring the nature of 
quantum measurement, we hope CCSN be part of a growing
interaction between the foundations of quantum mechanics and
gravitational physics, which may eventually lead to experimental
tests that shed light on both.

\acknowledgments

We thank Angelo Bassi, Mahdi Hosseini, Andrew A Geraci, Tabassom Hamidfar, Jinyang Li, Wisdom Boinde, Miguel Ceja, Mehrdad Sourki, and Vedant Kashyap for useful discussions. 
This work is supported by the Simons Foundation (Award No.\,568762), the Keck Foundation (Award No. AWD00002837), and the Army Research Office (Award No. W911NF202076).
D.M.\,is supported by the JSPS Overseas Research Fellowships.
Y.M.\,is supported by National Natural Science Foundation of China under Grants No. 12474481, No. 12441503.

\appendix
\begin{figure*}
    \includegraphics[width=0.8\textwidth]{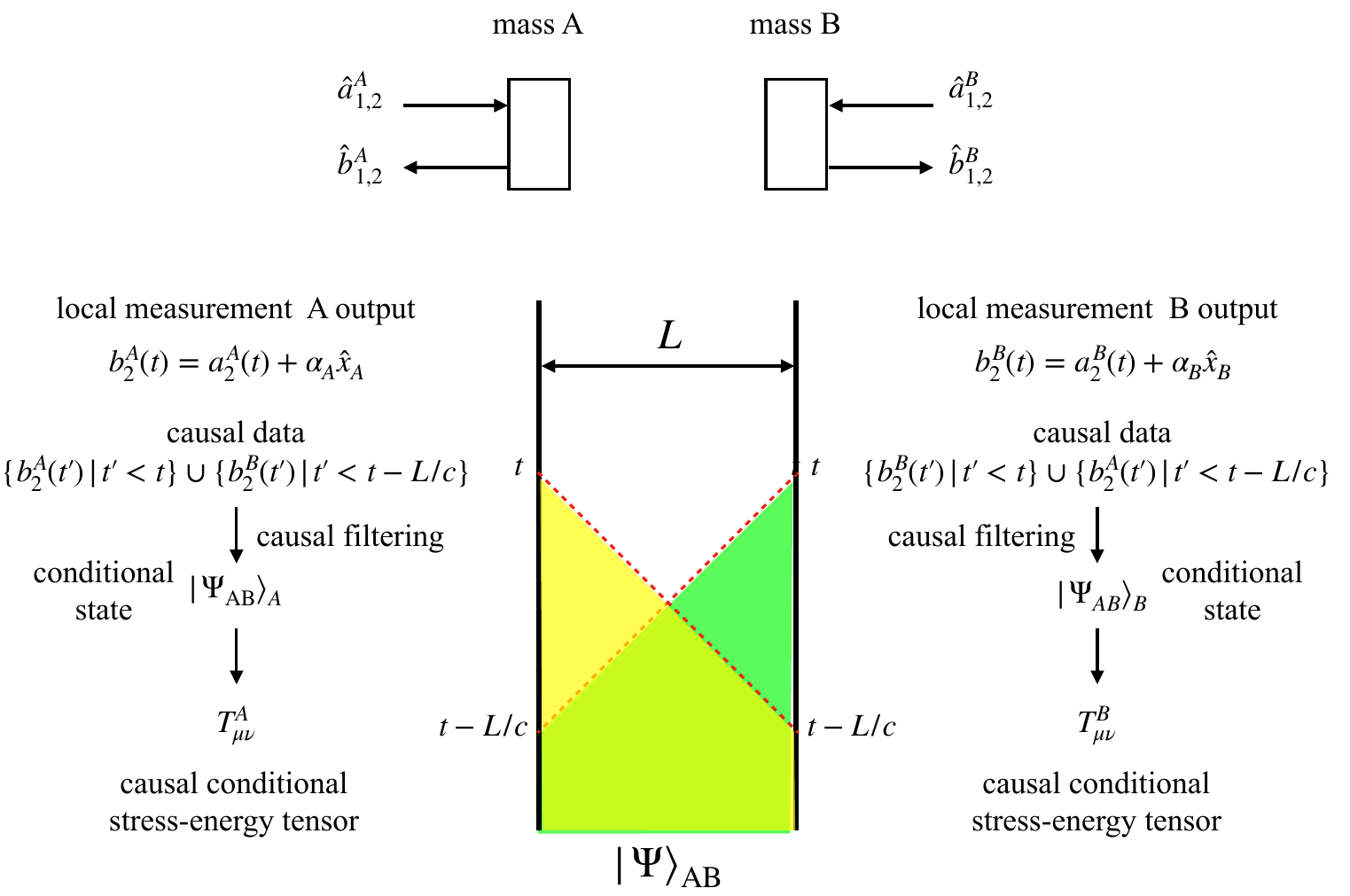}
    \caption{Causal Conditional Schr\"odinger-Newton theory in Heisenberg Picture. \label{fig:heisenerg}}
\end{figure*}

\section{Concrete realization of Kent's proposal/CCSN}

To be more concrete, let us consider objects $A$ and $B$, separated by distance $L$, each continuously monitored by an experimentalist, and explicitly write down what is necessary to form the local pure states at $A$ and $B$, and how to construct the feedback gravity.  To illustrate the causal structure, let us consider simple experiments that are performed as continuous linear measurements.  The way to model that is to have measurement strengths $\alpha_{A,B}$, incoming field quadratures $a_{1,2}^{A,B}$ and out-going field quadratures $b_{1,2}^{A,B}$. These quantities, plus the Heisenberg Operators of the positions of the masses, satisfy the following equations
\begin{align}
    D_A \hat x_A = \alpha_A a_1^A  + f_A,\; 
    b_1^A =a_1^A \,,\; b_2^A = a_2^A +\alpha_A \hat x_A \,,\\
        D_B \hat x_B = \alpha_B a_1^B  + f_B,\; 
    b_1^B =a_1^B \,,\; b_2^B = a_2^B +\alpha_B  \hat  x_B \,,
\end{align}
In these equations, each mass has its position operator  affected by the quantum radiation pressure force $\alpha_A a_1^A$ and the gravity force $f_A$, and the out-going phase quadratures $b_{2}^{A,B}$ are affected by the positions of the corresponding masses.  

Since we are in the Heisenberg picture, we only need to specify the initial states of the masses and the incoming fields, and the statistics of the out-going fields are completely specified by expectation values of the out-going operators. More specifically, each possible measurement record is a projection of the out-going field into simultaneous eigenstates of $\{ \hat b_2^A(t),\hat b_2^B(t)\}$ --- if we choose to measure the phase quadratures of the out-going fields. 

Feedback control is based on continuous measurements of $b_2^A$ and $b_2^B$, which yield results of $\xi^A$ and $\xi^B$. Explicitly, we need 
\begin{align}
    f_A(t) = \int_{-\infty}^t K_{AA}(t,t') \xi^A(t') dt'+
\int_{-\infty}^{t-L/c} K_{AB}(t,t') \xi^B(t') dt' \label{eqfA}\\
    f_B(t) = \int_{-\infty}^{t-L/c} K_{BA}(t,t') \xi^A(t') dt'+
\int_{-\infty}^{t} K_{BB}(t,t') \xi^B(t') dt'
\label{eqfB}
\end{align}
This is the most general treatment of causal feedback control.  However, in the case of causal conditional linearized gravity, the feedback is broken into two steps.  

The first step is to form a classical causal-conditional stress-energy tensor $T_{\mu\nu}$ associated with the masses, and then during the second step we form classical gravity from $T_{\mu\nu}$.  The causality of the second step is guaranteed by the Linearized Einstein's equation.  The first step is illustrated in Figure~\ref{fig:heisenerg}.  

Let us now lay down the precise steps. Basically, for measurement records $\xi^A(t)$ and $\xi^B(t)$, with $0\le t <+\infty$, we would like to obtain the probability density of obtaining such values. Given the classical forces $f_A$ and $f_B$ on $A$ and $B$ throughout history, we can write, in the Heisenberg Picture, 
\begin{align}
    \hat b_2^A (t) = \hat b_2^{A\,0} (t)  + \alpha_A \int_{0}^t \chi_A(t-t') f_A(t') dt' \\
        \hat b_2^B (t) = \hat b_2^{B\,0} (t)  + \alpha_B \int_{0}^t \chi_B(t-t') f_B(t') dt' \,,
\end{align}
where $\hat b_2^{A\,0} (t)$ and $\hat b_2^{B\,0} (t)$ are Heisenberg operators of the out-going field in absence of gravity, while $\chi_{A,B}$ are response functions of the individual masses. From this expression, we can obtain the probability density for obtaining a particular set of records:\begin{widetext}
\begin{align}
p[\xi^A,\xi^B] \propto \langle \Psi_{AB} | \mathbb{P}[\hat b_2^{A\,0}(t) =\xi^A(t)-\alpha_A\, (\chi_A \circ f_A)(t) \,,\; \hat b_2^{B\,0} (t) = \xi^B(t)-\alpha_B\, (\chi_B \circ f_B)(t) ]|\Psi_{AB}\rangle
\end{align}
\end{widetext}
Here $\mathbb{P}$ is a projection operator into the simultaneous eigenspace of $\hat b^{A0}_2$ and $\hat b^{B0}_2$ which all commute with each other. We have also used $\circ$ to denote convolution.   Within individual system, we have
\begin{equation}
[\hat b_2^A(t),\hat  b_2^A(t')] =[\hat b_2^B(t),\hat  b_2^B(t')]=0
    \end{equation}
since these are out-going fields, while across the two system these operators commute, with 
\begin{equation}
    [b_2^A(t) ,b_2^B(t')]=0
    \end{equation}
    since the two systems do not quantum mechanically interact with each other.  We will now only need to write down how $f_A$ and $f_B$ depend causally on $\xi_A$ and $\xi_B$ to complete a self-consistent set of equations. At time $t$, for each object, we will have to form a local classical value for the position of each object, which we shall label as  $x_c^A$ and $x_c^B$, respectively. If, for $A$, we are allowed to use A measurement data up to $t$ and $B$ measurement data up to $t-L/c$, we can formally use the {\it local pure states} of 
\begin{widetext}
\begin{align}
\label{localpurestates}
|\Psi_{AB}(t)\rangle_A  & =    \mathbb{P}\left[
\{\hat b_{2}^A(t') =\xi^A(t')|t'<t \} \,\& \,\{\hat b_{2}^B(t') =\xi^B(t')|t'<t-L/c \}  \right] |\psi_{AB}\rangle \\
|\Psi_{AB}(t)\rangle_B  & =    \mathbb{P}\left[
\{\hat b_{2}^A(t') =\xi^A(t')|t'<t-L/c \}\, \&\,\{\hat b_{2}^B(t') =\xi^B(t')|t'<t \}  \right] |\psi_{AB}\rangle
\end{align}
\end{widetext} From the local pure states, we can define 
\begin{align}
    T_A^{\mu\nu}(t,\mathbf{x}) = _A\langle \Psi_{AB}(t) |\hat T_A^{\mu\nu}|\Psi_{AB}(t)\rangle_A  \\
        T_B^{\mu\nu}(t,\mathbf{x}) = _B{}\langle  \Psi_{AB}(t) |\hat T_B^{\mu\nu}|\Psi_{AB}(t)\rangle_B 
        \label{TclassB}
\end{align}
where $T_A^{\mu\nu}$ only has support near the spatial location of $A$ and $T_B^{\mu\nu}$  only has support near the spatial location of $B$.  We will then have
\begin{equation}
   \bar h^{\mu\nu}(t,\mathbf{x}) = \int d^3\mathbf{x}'  \frac{T^{\mu\nu} (t-|\mathbf{x}-\mathbf{x}'|,\mathbf{x'})}{|\mathbf{x}-\mathbf{x}'|} 
\end{equation}
and obtain $f_A$ and $f_B$ from $\bar h^{\mu\nu}$.We note that from Eqs.~\eqref{localpurestates}--\eqref{TclassB},   $T_A^{\mu\nu}$ and $T_B^{\mu\nu}$ are computed classically from the causal data and the initial state $|\Psi_{AB}\rangle$.  Since initial state can be inferred from earlier measurements and setup of the experiment, this can be viewed as classical knowledge as well. In this way, the gravity field is generated in a causal way directly from experimental data, and implemented onto each mass via a local Hamiltonian.  More specifically, we recover dependence of the form \eqref{eqfA} and \eqref{eqfB}.

In the limit when time of a single object, or when light travel time between the two objects are negligible, we recover the CCSN formulation obtained by Liu et al.~\cite{liu2023semiclassical}.

\section{Corrector field from a causally delayed conditional-state update}

The continuous-measurement example considered in Sec.~IV D describes the corrector field generated by the stochastic conditional evolution of a continuously monitored self-gravitating test mass. We now consider a complementary situation based directly on the thought experiment of Sec.~II B. We retain the two systems $A$ and $B$ in order to make the causal propagation of the measurement record explicit, while focusing on the self-gravity of $B$.

Consider again the entangled state in Eq.~(6), where the two systems are separated by a distance $L$. A localized measurement is performed on $A$ around time $t_M$. As discussed in Sec.~II B, for an ideal projective measurement the outcome is immediately available locally at $A$, whereas the local conditional state used to source gravity near $B$ does not respond until the corresponding classical record enters its past light cone, at approximately $t_B=t_M+L/c$. Thus, the causal delay considered here is the propagation time of the measurement record from $A$ to $B$. In a realistic measurement, the measurement record is generated over a finite time. We denote by $\tau_M$ the characteristic duration over which the measurement on $A$ resolves the outcome and the corresponding conditional state changes appreciably from its prior to its posterior state. The ideal projective-measurement limit corresponds to $\tau_M\rightarrow0$.

To describe this causal update, we introduce a switching function $f_{\tau_M}\left(t-t_M-\frac{|\mathbf{x}-\mathbf{x}_A|}{c}\right)$, which changes from zero to unity over a time scale of order $\tau_M$. The conditional stress-energy tensor sourcing the self-gravity of $B$ is then written schematically as
\begin{equation}
    T_{B,\rm SN}^{\mu\nu}(x)=
    T_{B,\rm pri}^{\mu\nu}(x)
    +f_{\tau_M}\left(t-t_M-\frac{|\mathbf{x}-\mathbf{x}_A|}{c}\right)
    \Delta T_B^{\mu\nu}(x),
\end{equation}
where $\Delta T_B^{\mu\nu}=T_{B,\rm post}^{\mu\nu}-T_{B,\rm pri}^{\mu\nu}$. If the prior and posterior stress-energy tensors are separately conserved away from the transition region, the measurement-induced non-conservation is
\begin{equation}
    J_B^\mu=
    \nabla_\nu T_{B,\rm SN}^{\mu\nu}
    \simeq
    \nabla_\nu
    f_{\tau_M}\left(t-t_M-\frac{|\mathbf{x}-\mathbf{x}_A|}{c}\right)
    \Delta T_B^{\mu\nu}.
\end{equation}
Thus, the corrector field is required because the causal measurement-induced update makes the conditional stress-energy tensor locally non-conserved during the transition. Equivalently, on a constant-time slice intersecting the causal transition region, the spatial integral of the conditional energy density need not remain equal to its prior or posterior value, because different spatial points are conditioned on different amounts of the measurement record. This temporary mismatch is an integrated manifestation of $J_B^0\neq0$ and illustrates why an additional corrector contribution is required to restore energy-momentum conservation.

In the ideal instantaneous-measurement limit, the non-conservation current becomes distributionally localized on the causal update front. The corresponding corrector field is obtained from Eqs.~(73)--(76), with the retarded boundary condition adopted in Sec.~IV C. Since $J_B^\mu$ is generated during the causal transition, the resulting corrector has a transient retarded character associated with the conditional-state update of $B$.

The ordinary CCSN self-gravity and the corrector field have qualitatively different dependences on this causal update. The measurement-induced change in the conditional state changes the Schr\"odinger--Newton potential by
\begin{equation}
    \Delta\Phi_{B,\rm SN}=
    -G\int d^3x'\,
    \frac{\Delta\rho_B(\mathbf{x}')}{|\mathbf{x}-\mathbf{x}'|},
\end{equation}
where $\Delta\rho_B=\rho_{B,\rm post}-\rho_{B,\rm pri}$. This change modifies the subsequent self-gravitational evolution of $B$ after the measurement record has arrived. By contrast, the corrector is sourced by $J_B^\mu$, and hence by spacetime derivatives of the switching function. The ordinary CCSN response is therefore associated with the conditional-state update itself, while the corrector response is generated primarily during the transition.

Schematically, a gravitational or mechanical readout sensitive to the field sourced by $B$ may therefore be written as
\begin{equation}
    y(t)=
    y_{\rm pri}(t)
    +y_{\rm SN}[f_{\tau_M}](t)
    +y_{\rm corr}[f_{\tau_M}](t),
\end{equation}
where $y_{\rm SN}[f_{\tau_M}](t)$ denotes the ordinary CCSN response associated with the causally updated conditional state, whereas $y_{\rm corr}[f_{\tau_M}](t)$ denotes the additional response generated by the corrector field. The latter depends on the same switching function through $J_B^\mu$, and in particular through its spacetime derivatives. If a separate readout probe is located a distance $R$ from $B$, the relevant response is further delayed by the propagation time from $B$ to the probe. Both contributions are therefore causally tied to the measurement on $A$, but they have different temporal structures.

This difference makes the corrector contribution, at least in principle, experimentally identifiable. The measurement time $t_M$, the separation $L$, and the measurement protocol are known experimentally. Consequently, the corrector response is not an unconstrained stochastic excess: its temporal structure is determined by the same causal measurement update that generates $J_B^\mu$. Experimentally testing the corrector therefore amounts to comparing the ordinary CCSN prediction with the prediction obtained after adding the corrector contribution. A quantitative search may be formulated as a noise-weighted search for this additional transient response, or equivalently as a likelihood comparison between models with and without the corrector field.

\bibliography{main}

\end{document}